\documentstyle[multicol,aps,prl]{revtex} 

\renewcommand{\narrowtext}{\begin{multicols}{2} \global\columnwidth20.5pc}
\renewcommand{\widetext}{\end{multicols} \global\columnwidth42.5pc}
\def\top#1{\vskip #1\begin{picture}(290,80)(80,500)\thinlines \put(
65,500){\line( 1, 0){255}}\put(320,500){\line( 0, 1)
{ 5}}\end{picture}}
\def\bottom#1{\vskip #1\begin{picture}(290,80)(80,500)\thinlines \put(
330,500){\line( 1, 0){255}}\put(330,500){\line( 0, -1){
5}}\end{picture}}

\def\al{\alpha}
\def\be{\beta}
\def\ga{\gamma}
\def\de{\delta}
\def\ep{\epsilon}

\def\ze{\zeta}
\def\et{\eta}
\def\th{\theta}

\def\ka{\kappa}
\def\la{\lambda}

\def\rh{\rho}

\def\si{\sigma}

\def\ta{\tau}

\def\ph{\phi}

\def\ch{\chi}
\def\ps{\psi}
\def\om{\omega}
\def\Ga{\Gamma}
\def\De{\Delta}
\def\Th{\Theta}
\def\La{\Lambda}

\def\mn{{\mu\nu}}

\def\cl{{\cal L}}
\def\cE{{\cal E}}
\def\pt#1{\phantom{#1}}
\def\prt{\partial}
\def\vev#1{\langle {#1}\rangle}

\def\fr#1#2{{{#1} \over {#2}}}
\def\frac#1#2{{\textstyle{{#1}\over {#2}}}}
\def\half{{\textstyle{1\over 2}}}
\def\lsim{\mathrel{\rlap{\lower4pt\hbox{\hskip1pt$\sim$}}
    \raise1pt\hbox{$<$}}}
\def\gsim{\mathrel{\rlap{\lower4pt\hbox{\hskip1pt$\sim$}}
    \raise1pt\hbox{$>$}}}
\def\sqr#1#2{{\vcenter{\vbox{\hrule height.#2pt
         \hbox{\vrule width.#2pt height#1pt \kern#1pt
         \vrule width.#2pt}
         \hrule height.#2pt}}}}

\def\lrprtmu{\stackrel{\leftrightarrow}{\partial_\mu}}

\def\lrDmu{\stackrel{\leftrightarrow}{D_\mu}}
\def\lrDnu{\stackrel{\leftrightarrow}{D^\nu}}

\def\Re{\hbox{Re}\,}

\newcommand{\beq}{\begin{equation}}
\newcommand{\eeq}{\end{equation}}
\newcommand{\bea}{\begin{eqnarray}}
\newcommand{\eea}{\end{eqnarray}}
\newcommand{\rf}[1]{(\ref{#1})}
 
\begin{document}

\title{Lorentz-Violating Extension of the Standard Model}    
\author{D.\ Colladay and V.\ Alan Kosteleck\'y}
\address{Physics Department, Indiana University, 
          Bloomington, IN 47405, U.S.A.}
\date{preprint IUHET 359 (1997); 
accepted for publication in Phys.\ Rev.\ D}
\maketitle

\begin{abstract}
In the context of conventional quantum field theory,
we present a general Lorentz-violating extension
of the minimal SU(3) $\times$ SU(2)$ \times$ U(1) standard model 
including CPT-even and CPT-odd terms.
It can be viewed as the low-energy limit of a physically relevant 
fundamental theory with Lorentz-covariant dynamics 
in which spontaneous Lorentz violation occurs.
The extension has gauge invariance,
energy-momentum conservation,
and covariance under observer rotations and boosts,
while covariance under particle rotations and boosts is broken. 
The quantized theory is hermitian and power-counting renormalizable,
and other desirable features such as
microcausality, 
positivity of the energy,
and the usual anomaly cancellation are expected.
Spontaneous symmetry breaking to the electromagnetic U(1) 
is maintained,
although the Higgs expectation is shifted by a small amount 
relative to its usual value
and the $Z^0$ field acquires a small expectation.
A general Lorentz-breaking extension 
of quantum electrodynamics is extracted from the theory,
and some experimental tests are considered.
In particular,
we study modifications to photon behavior. 
One possible effect is vacuum birefringence,
which could be bounded from cosmological observations
by experiments using existing techniques.
Radiative corrections to the photon propagator are examined.
They are compatible with
spontaneous Lorentz and CPT violation in the fermion sector
at levels suggested by Planck-scale physics
and accessible to other terrestrial laboratory experiments.

\end{abstract}


\narrowtext

\section{INTRODUCTION}

The minimal SU(3)$\times$SU(2)$\times$U(1) standard model,
although phenomenologically successful,
leaves unresolved a variety of issues.
It is believed to be the low-energy limit of a fundamental theory
that also provides a quantum description of gravitation.
An interesting question is whether any aspects of 
this underlying theory could be revealed 
through definite experimental signals
accessible with present techniques.

The natural scale for a fundamental theory including gravity 
is governed by the Planck mass $M_P$,
which is about 17 orders of magnitude greater than
the electroweak scale $m_W$ associated with the standard model.
This suggests that observable experimental signals
from a fundamental theory 
might be expected to be suppressed 
by some power of the ratio 
$r \approx m_W/M_P \simeq 10^{-17}$.
Detection of these minuscule effects at present energy scales
would be likely to require experiments of exceptional sensitivity,
preferably ones seeking to observe a signal forbidden 
in conventional renormalizable gauge theories.

To identify signals of this type,
one approach is to examine proposed fundamental theories
for effects that are qualitatively different 
from standard-model physics.
For example,
at present the most promising framework for a fundamental theory
is string (M) theory.
The qualitative difference between particles and strings
means that qualitatively new physics is expected at the Planck scale.
An interesting challenge would be to determine whether
this could lead to observable low-energy effects. 

In the present work,
we consider the possibility that the new physics
involves a violation of Lorentz symmetry.
It has been shown that spontaneous Lorentz breaking
may occur in the context of string theories
with Lorentz-covariant dynamics
\cite{kps}.
Unlike the conventional standard model,
string theories typically involve interactions 
that could destabilize the naive vacuum
and trigger the generation of
nonzero expectation values for Lorentz tensors.
Note that some kind of spontaneous breaking of 
the higher-dimensional Lorentz symmetry is expected 
in any realistic Lorentz-covariant fundamental theory 
involving more than four spacetime dimensions.
If the breaking extends 
into the four macroscopic spacetime dimensions,
apparent Lorentz violation could occur 
at the level of the standard model.
This would represent a possible observable effect
from the fundamental theory,
originating outside the structure 
of conventional renormalizable gauge models.

A framework has been developed for treating the effects 
of spontaneous Lorentz breaking  
in the context of a low-energy effective theory
\cite{cksm},
where certain terms can be induced that appear 
to violate Lorentz invariance explicitly.
It turns out that,
from a theoretical perspective,
the resulting effects 
are comparatively minimal.

An important point is that
Lorentz symmetry remains a property
of the underlying fundamental theory
because the breaking is spontaneous.
This implies that various attractive features 
of conventional theories,
including microcausality and positivity of the energy,
are expected to hold 
in the low-energy effective theory.
Also,
energy and momentum are conserved as usual,
provided the tensor expectation values in the fundamental theory
are spacetime-position independent.
Moreover,
standard quantization methods are unaffected,
so a relativistic Dirac equation
and a nonrelativistic Schr\"odinger equation
emerge in the appropriate limits.

Another important aspect of the spontaneous breaking
is that both the fundamental theory and the effective
low-energy theory remain invariant under
\it observer \rm Lorentz transformations,
i.e., rotations or boosts of an observer's inertial frame
\cite{cksm}. 
The presence of nonzero tensor expectation values
in the vacuum affects only invariance properties 
under \it particle \rm Lorentz transformations,
i.e., rotations or boosts of a localized particle or field
that leave unchanged the background expectation values.

This framework for treating spontaneous Lorentz violation
has been used to obtain a general extension 
of the minimal SU(3) $\times$ SU(2)$ \times$ U(1) standard model
that violates both Lorentz invariance and CPT
\cite{cksm}.
In addition to the desirable features of 
energy-momentum conservation,
observer Lorentz invariance,
conventional quantization,
hermiticity,
and the expected microcausality and positivity
of the energy,
this standard-model extension maintains gauge invariance 
and power-counting renormalizability.
It would emerge from any fundamental theory 
(not necessarily string theory) 
that generates the standard model
and contains spontaneous Lorentz and CPT violation.

The present work continues 
our previous theoretical investigations
of spontaneous Lorentz and CPT breaking.
Working first at the level of the standard model,
we provide explicitly in section II
the full Lorentz-violating extension,
including the CPT-even Lorentz-breaking terms
described implicitly in Ref.\ \cite{cksm}.
We also give some details of the modifications
to the usual electroweak symmetry breaking.

Since many sensitive measures of Lorentz and CPT symmetry
involve tests of quantum electrodynamics (QED),
it is useful to extract from the standard-model extension
a generalized QED 
that allows for possible Lorentz and CPT violations.
This extended QED, 
given in section III,
involves modifications of the usual QED
in both the fermion and the photon sectors.
Some comments are also given in section III 
about the implications of this theory
for experimental tests with electrons and positrons.

In the remainder of this paper, 
we focus primarily on the photon sector of the extended QED,
presenting a study of the theoretical and experimental 
implications of the modifications to photon properties
arising from the possible Lorentz and CPT violations.
Section IV discusses changes in the basic theory,
including the modified Maxwell equations
and properties of their solutions.
One possible effect is vacuum photon birefringence,
and some associated features are described.
We show that feasible measurements limiting birefringence 
on cosmological scales could tightly constrain
the Lorentz-violating terms.
In section V,
some important consistency checks on the theory
at the level of radiative corrections are presented,
largely at the one-loop level.
The types of Lorentz violation
that can be affected by radiative corrections
are identified,
and explicit calculations are given.
We show that the effects are compatible with
spontaneous Lorentz and CPT violation in the fermion sector
at levels accessible to other QED experiments. 

Since the standard-model extension
provides a quantitative microscopic theory of Lorentz and CPT violation,
it is feasible to identify potentially observable signals 
and to establish bounds from various experiments
other than ones in the photon sector.
Numerous tests of Lorentz invariance and CPT exist.
The present theory provides a single coherent framework
at the level of the standard model and QED
that can be used as a basis for the analysis and comparison
of these tests.
Although many experiments are insensitive to 
the suppressed effects motivating our investigation,
certain high-precision ones might have observable signals
within this framework.
In particular,
the results in the present paper have been used
to examine possible bounds on CPT and Lorentz violations 
from measurements of neutral-meson oscillations
\cite{ckpv,kexpt,bexpt,ak},
from tests of QED 
in Penning traps
\cite{pennexpts,bkr},
and from baryogenesis
\cite{bckp}.
Several other investigations are underway,
including a study 
\cite{bkr2}
of possible Lorentz and CPT effects
on hydrogen and antihydrogen spectroscopy
\cite{antih}
and another 
\cite{kla}
of limits attainable in
clock-comparison experiments
\cite{cc}.

The analyses of the standard-model and QED extensions
performed in the present work
leave unaddressed a number of significant theoretical issues
arising at scales between the electroweak mass
and the Planck mass.
These include the `dimension problem'
of establishing whether spontaneous Lorentz breaking 
in the fundamental theory near the Planck scale
indeed extends to the four physical spacetime dimensions
and, if so,
the mechanism for its suppression
or, if not,
why exactly four spacetime dimensions are spared.
Other issues include the effects of mode fluctuations
around the tensor expectation values 
and possible constraints and effects arising from 
a nonminimal standard model or
(super)unification below the Planck scale.  

Another potentially important topic is the implication
of spontaneous Lorentz violation for gravity at observable energies.
Like the usual standard model,
the standard-model extension 
considered here disregards gravitational effects.
The particle Lorentz symmetry 
that is broken in this theory 
is therefore effectively a global symmetry,
and so one might expect Nambu-Goldstone modes.
Since gravity is associated with local Lorentz invariance,
it is natural to ask about the role of these modes 
in a version of the standard-model extension that includes gravity.
In a gauge theory,
when a suitable scalar acquires a nonzero expectation value,
the Higgs mechanism occurs:
the propagator for the gauge boson is modified,
and a mass is generated. 
Similarly,
in a theory with gravitational couplings,
when a Lorentz tensor acquires a nonzero expectation value,
the graviton propagator can be modified.
However,
no mass for the graviton is induced 
because the gravitational connection 
is related to the derivative of the metric
rather than to the metric itself
\cite{kps}.
In this sense,
there is no gravitational Higgs effect.

The theory described here 
appears at present to be the sole candidate
for a consistent extension of the standard model
providing a microscopic theory of Lorentz violation.
A complete review of alternative approaches 
to possible Lorentz and CPT violation
lies beyond the scope of this paper. 
Works known to us of relevance in the present context 
are referenced in the body of the text below.
Among other ideas in the literature are several distinctive ones 
developed from perspectives very different from ours.
Following early work by Dirac and Heisenberg,
several authors have considered 
an unphysical spontaneous Lorentz breaking in an effort 
to interpret the photon as a Nambu-Goldstone boson
\cite{dh}.
Nielsen and his colleagues have suggested
the converse of the philosophy in the present work:
that the observed Lorentz symmetry in nature
might be a low-energy manifestation
of a fundamental theory \it without \rm Lorentz invariance.
A discussion of this idea 
and a brief review of the literature on Lorentz breaking 
prior to the establishment of the usual minimal standard model
may be found in Ref.\ \cite{np}.
Hawking has suggested 
\cite{swh}
the possibility that conventional quantum mechanics
is invalidated by gravitational effects
and that this might lead to CPT violation,
among other effects. 
The implications for experiments in the kaon system
\cite{elmn}
are known to be entirely different from
those arising in the present standard-model extension,
which is based on conventional quantum theory.
There is also a body of literature pertaining
to unconventional theories of gravity
(without standard-model physics),
among which are some models containing various 
possible sources of local Lorentz violation
\cite{cw}.

\section{STANDARD-MODEL EXTENSION}

In this section,
we extend the minimal standard model
by adding all possible Lorentz-violating terms
that could arise from spontaneous symmetry breaking
at a fundamental level
but that preserve
SU(3) $\times$ SU(2)$ \times$ U(1) 
gauge invariance 
and power-counting renormalizability.
Terms that are odd under CPT
are explicitly given in Ref.\ \cite{cksm}
but are also included here for completeness.

The general form of a Lorentz-violating term
involves a part that acts as a coupling coefficient and 
a part constructed from the basic fields in the standard model.
The requirements of the derivation impose various limitations 
on the possible structures of both parts.
Taken together,
these requirements place significant constraints
on the form of terms in the standard-model extension. 

The part acting as a coupling coefficient 
carries spacetime indices reflecting the properties 
under observer Lorentz transformations
of the relevant nonzero expectation values 
from the fundamental theory.
The coupling coefficient may be complex,
but it is constrained by 
the requirement that the lagrangian be hermitian.
For a coupling coefficient with an even number of spacetime indices,
the pure trace component is irrelevant for present purposes
because it maintains Lorentz invariance.
A coupling coefficient of this type can therefore be taken traceless.

The field part may involve covariant derivatives and,
if fermions are involved,
gamma matrices.
Gauge invariance requires 
that the field part be a singlet under 
SU(3) $\times$ SU(2)$ \times$ U(1),
while power-counting renormalizability implies that
it must have mass dimension no greater than four.
The requirement that the standard-model extension  
originates from spontaneous Lorentz breaking 
in a covariant fundamental theory 
implies the whole Lorentz-violating term
must be a singlet under observer Lorentz transformations,
so the field part must have indices
matching those of the coupling coefficient.

Following the discussion in the introduction,
all coupling coefficients are assumed to be heavily suppressed
by some power of the ratio $r$ 
of the light scale to the Planck scale.
In the absence of a satisfactory explanation of
the suppression mechanism,
it would seem premature to attempt specific detailed predictions 
about the relative sizes of different coupling coefficients. 
As a possible working hypothesis,
one might attribute comparable suppression factors to all terms 
at the level of the standard-model extension.
Note that a term with field part having mass dimension $n$ 
must have a coupling coefficient with mass dimension $4-n$,
and the relevant scale for these effects 
is roughly the Planck mass.
The hypothesis would therefore suggest that 
in the low-energy theory 
a term with field part of mass dimension $n+1$ 
would have coupling coefficient 
suppressed by an additional power of $r$
relative to the coefficient of a field term 
of mass dimension $n$.
This scheme would be compatible with interpreting
the standard model as an effective field theory,
in which each additional derivative coupling would 
involve an additional suppression factor
in the coupling coefficient.
It would imply a distinct hierarchy among
the coupling coefficients introduced below,
and would suggest that certain derivative couplings
could be neglected relative to comparable nonderivative ones.
However, 
since this hypothesis presently has no basis 
in a detailed theory,
in what follows we have chosen to retain on an equal footing
all renormalizable terms compatible with
the gauge symmetries of the standard model
and with an origin in spontaneous Lorentz breaking. 

In what follows,
we denote the left- and right-handed 
lepton and quark multiplets by
$$
L_A = \left( \begin{array}{c} \nu_A \\ l_A
\end{array} \right)_L
\quad , \quad
R_A = (l_A)_R
\quad ,
$$
\beq
Q_A = \left( \begin{array}{c} u_A \\ 
d_A \end{array} \right)_L ~~ , ~~ 
U_A = (u_A)_R ~~ , ~~ D_A = (d_A)_R
\quad , 
\eeq
where
\beq
\ps_L \equiv \frac 1 2 ( 1 - \ga_5 ) \ps
\quad , \qquad
\ps_R \equiv \frac 1 2 ( 1 + \ga_5 ) \ps
\quad ,
\label{handproj}
\eeq
as usual,
and where $A = 1,2,3$ labels the flavor:
$l_A \equiv (e, \mu, \ta)$, 
$\nu_A \equiv (\nu_e, \nu_\mu, \nu_\ta)$,
$u_A \equiv (u,c,t)$, $d_A \equiv (d,s,b)$.
We denote the Higgs doublet by $\ph$,
and in unitary gauge we represent it as 
\beq
\ph = \fr 1 {\sqrt{2}} 
\left( \begin{array}{c} 0 \\ r_\ph 
\end{array} \right)
\quad .
\eeq
The conjugate doublet is $\ph^c$.
The SU(3), SU(2), and U(1) gauge fields
are denoted by
$G_\mu$, $W_\mu$, and $B_\mu$, respectively.
The corresponding field strengths are 
$G_{\mu\nu}$, $W_{\mu\nu}$, and $B_{\mu\nu}$, 
with the first two understood to be
hermitian adjoint matrices
while $B_{\mu\nu}$ is a hermitian singlet.
The corresponding couplings
are $g_3$, $g$, and $g^\prime$.
The electromagnetic U(1) charge $q$ and the angle $\th_W$
are defined through $q = g \sin \th_W = g^\prime \cos \th_W$,
as usual.
The covariant derivative is denoted by $D_\mu$,
and $A\lrprtmu B \equiv A\prt_\mu B - (\prt_\mu A) B$.
The Yukawa couplings are $G_L$, $G_U$, $G_D$.
Throughout most of this work we use natural units,
which could be obtained from the SI system 
by redefining $\hbar = c = \ep_0 = 1$,
and we adopt the Minkowski metric $\et_{\mu\nu}$
with $\et_{00}= +1$.

The complete lagrangian for the Lorentz-breaking
standard-model extension can be separated into
a sum of terms.
For completeness,
we first provide the lagrangian terms in the 
usual SU(3) $\times$ SU(2) $\times$ U(1) 
minimal standard model:
\beq
\cl_{\rm lepton} = 
\half i \overline{L}_A \ga^{\mu} \lrDmu L_A
+ \half i \overline{R}_A \ga^{\mu} \lrDmu R_A
\label{smlepton}
\quad ,
\eeq
\bea
\cl_{\rm quark} &=&
\half i \overline{Q}_A \ga^{\mu} \lrDmu Q_A
+ \half i \overline{U}_A \ga^{\mu} \lrDmu U_A
\nonumber \\ &&
+ \half i \overline{D}_A \ga^{\mu} \lrDmu D_A
\quad ,
\label{smquark}
\eea
\bea
\cl_{\rm Yukawa} &=& 
- \left[ (G_L)_{AB} \overline{L}_A \ph R_B
+ (G_U)_{AB} \overline{Q}_A \ph^c U_B 
\right.
\nonumber \\ &&
\left.
\qquad + (G_D)_{AB} \overline{Q}_A \ph D_B
\right]
+ {\rm h.c.}
\quad ,
\label{smyukawa}
\eea
\beq
\cl_{\rm Higgs} =
(D_\mu\ph)^\dagger D^\mu\ph 
+\mu^2 \ph^\dagger\ph - \fr \la {3!} (\ph^\dagger\ph)^2
\quad ,
\label{smhiggs}
\eeq
\beq
\cl_{\rm gauge} =
-\half {\rm Tr} (G_{\mu\nu}G^{\mu\nu})
-\half {\rm Tr} (W_{\mu\nu}W^{\mu\nu})
-\frac 1 4 B_{\mu\nu}B^{\mu\nu} ~ .
\label{smgauge}
\eeq
The usual $\th$ terms have been omitted,
and possible analogous total-derivative terms 
that break Lorentz symmetry
are disregarded in this work.

In the fermion sector of the standard-model extension,
the contribution to the lagrangian can be divided into four parts
according to whether the term is CPT even or odd
and whether it involves leptons or quarks:
\bea
\cl^{\rm CPT-even}_{\rm lepton} &=& 
\half i (c_L)_{\mu\nu AB} \overline{L}_A \ga^{\mu} \lrDnu L_B
\nonumber\\ &&
+ \half i (c_R)_{\mu\nu AB} \overline{R}_A \ga^{\mu} \lrDnu R_B
\label{lorvioll}
\quad ,
\eea
\beq
\cl^{\rm CPT-odd}_{\rm lepton} = 
- (a_L)_{\mu AB} \overline{L}_A \ga^{\mu} L_B
- (a_R)_{\mu AB} \overline{R}_A \ga^{\mu} R_B
\label{cptvioll}
\quad ,
\eeq
\bea
\cl^{\rm CPT-even}_{\rm quark} &=& 
\half i (c_Q)_{\mu\nu AB} \overline{Q}_A \ga^{\mu} \lrDnu Q_B
\nonumber\\ &&
+ \half i (c_U)_{\mu\nu AB} \overline{U}_A \ga^{\mu} \lrDnu U_B 
\nonumber\\ &&
+ \half i (c_D)_{\mu\nu AB} \overline{D}_A \ga^{\mu} \lrDnu D_B
\quad ,
\label{lorviolq}
\eea
\bea
\cl^{\rm CPT-odd}_{\rm quark} &=& 
- (a_Q)_{\mu AB} \overline{Q}_A \ga^{\mu} Q_B
- (a_U)_{\mu AB} \overline{U}_A \ga^{\mu} U_B 
\nonumber\\ &&
- (a_D)_{\mu AB} \overline{D}_A \ga^{\mu} D_B
\quad .
\label{cptviolq}
\eea
In these equations,
the various coupling coefficients $c_{\mu\nu}$ and $a_\mu$ 
are understood to be hermitian in generation space.
The coefficients $a_\mu$ have dimensions of mass.
The dimensionless coefficients $c_{\mu\nu}$ 
can have both symmetric and antisymmetric parts
but can be assumed traceless.
A nonzero trace would not contribute to Lorentz violation
and in any case can 
be absorbed by a conventional field normalization
ensuring the usual kinetic operator for the matter fields.

The standard-model extension also contains 
Lorentz-violating couplings
between the fermions and the Higgs field,
having the gauge structure of the usual Yukawa couplings 
but involving nontrivial gamma matrices.
These terms are all CPT even:
\bea
\cl^{\rm CPT-even}_{\rm Yukawa} = 
- \half &&
\left[
(H_L)_{\mu\nu AB} \overline{L}_A \ph \si^{\mu\nu} R_B
\right.
\nonumber\\ &&
\left.
+(H_U)_{\mu\nu AB} \overline{Q}_A \ph^c \si^{\mu\nu} U_B 
\right.
\nonumber\\ &&
\left.
+(H_D)_{\mu\nu AB} \overline{Q}_A \ph \si^{\mu\nu} D_B
\right]
+ {\rm h.c.}
\quad
\label{loryukawa}
\eea
The coefficients $H_{\mu\nu}$ are 
dimensionless and antisymmetric,
but like the Yukawa couplings $G_{L,U,D}$ 
they are \it not \rm necessarily hermitian in generation space.

The possible contributions in the Higgs sector
can be CPT even or CPT odd: 
\bea
\cl^{\rm CPT-even}_{\rm Higgs} &=&
\half (k_{\ph\ph})^{\mu\nu} (D_\mu\ph)^\dagger D_\nu\ph 
+ {\rm h.c.}
\nonumber\\ &&
-\half (k_{\ph B})^{\mu\nu} \ph^\dagger \ph B_{\mu\nu}
\nonumber\\ &&
-\half (k_{\ph W})^{\mu\nu} \ph^\dagger W_{\mu\nu} \ph 
\quad ,
\label{lorhiggs}
\eea
\beq
\cl^{\rm CPT-odd}_{\rm Higgs}
= i (k_\ph)^{\mu} \ph^{\dagger} D_{\mu} \ph + {\rm h.c.} 
\quad .
\label{cpthiggs}
\eeq
In Eq.\ \rf{lorhiggs},
the dimensionless coefficient $k_{\ph\ph}$ 
can have symmetric real
and antisymmetric imaginary parts.
The other coefficients in Eq.\ \rf{lorhiggs}
have dimensions of mass and must be real antisymmetric.
The coefficient
$k_\ph$ for the CPT-odd term \rf{cpthiggs}
also has dimensions of mass 
and can be an arbitrary complex number.

The gauge sector has both
CPT-even and CPT-odd contributions.
The CPT-even ones are:
\bea
\cl^{\rm CPT-even}_{\rm gauge} &=&
-\half (k_G)_{\ka\la\mu\nu} {\rm Tr} (G^{\ka\la}G^{\mu\nu})
\nonumber\\ &&
-\half (k_W)_{\ka\la\mu\nu} {\rm Tr} (W^{\ka\la}W^{\mu\nu})
\nonumber\\ &&
-\frac 1 4 (k_B)_{\ka\la\mu\nu} B^{\ka\la}B^{\mu\nu}
\quad .
\label{lorgauge}
\eea
In this equation,
the dimensionless coefficients $k_{G,W,B}$ are real.
They must have the symmetries of the Riemann tensor
and a vanishing double trace.
The point is that any totally antisymmetric part 
involves only a total derivative in the lagrangian density,
while a nonzero double trace can be absorbed
into a redefinition of the normalization 
of the corresponding kinetic term \rf{smgauge}.

The CPT-odd gauge terms are given by the following expression
\cite{fn1}:
\bea
\cl^{\rm CPT-odd}_{\rm gauge} &=&
(k_3)_\ka \ep^{\ka\la\mu\nu} 
{\rm Tr} (G_\la G_{\mu\nu} + \frac 2 3 i g_3 G_\la G_\mu G_\nu)
\nonumber\\ &&
+ (k_2)_\ka \ep^{\ka\la\mu\nu} 
{\rm Tr} (W_\la W_{\mu\nu} + \frac 2 3 i g W_\la W_\mu W_\nu)
\nonumber\\ &&
+ (k_1)_\ka \ep^{\ka\la\mu\nu} B_\la B_{\mu\nu} 
+ (k_0)_\ka B^\ka 
\quad .
\label{cptgauge}
\eea
The coefficients $k_{1,2,3}$ are real and have dimensions of mass,
while $k_0$ is also real and has dimensions of mass cubed.
It turns out that,
if any of these CPT-odd terms do indeed appear,
they would generate instabilities in the minimal theory.
They are all associated with negative 
contributions to the energy,
and in addition the term with $k_0$ would directly generate 
a linear instability in the potential.
It might therefore seem desirable 
that all the coefficients $k_{0,1,2,3}$ vanish.
While this could be imposed at the classical level,
radiative quantum corrections from, say, the fermion sector
might \it a priori \rm be expected to generate nonzero values.
Remarkably,
the structure of the standard-model extension appears to be
such that no corrections arise,
at least to one loop.
These issues are discussed further in what follows,
in particular in subsection IV A and section V.

It is known that some apparently CPT- and Lorentz-violating terms
can be eliminated from the action via field redefinitions
\cite{cksm}.
Several types of redefinition can be considered.
In the context of the present standard-model extension,
we have investigated a variety of possibilities for each field.
As a general rule,
the more complex the theoretical structure becomes,
the less likely it is that a useful field redefinition exists.
For instance,
the presence of Lorentz-violating CPT-even derivative couplings 
in the standard-model extension complicates 
the analysis for CPT-odd terms provided in Ref.\ \cite{cksm},
although it turns out that the conclusions still hold.
Here,
we summarize a few methodological results 
and describe some special cases of particular interest. 

To eliminate a Lorentz-breaking term,
a field redefinition must involve 
the associated coupling coefficient.
When derivative couplings play a role,
the field redefinition may also involve 
spacetime-position variables.
The assumption that the coupling coefficients are small
can be helpful,
in some cases directly assisting in derivations 
and in others leading to a set of 
approximate field redefinitions.
Under the latter, 
a theory with first-order Lorentz-breaking effects
may be redefined into one with effects 
appearing only at second or higher orders.
Alternatively,
some first-order Lorentz-breaking terms
may be absorbed into others.
A partial constraint on allowable redefinitions
is provided by the transformation properties of
the various Lorentz-violating terms 
under the discrete symmetries C, P, T.
Only terms with identical discrete-symmetry properties
can be absorbed into one another by first-order redefinitions.

Two types of redefinition that we have found of particular value
are linear phase redefinitions
and linear normalization redefinitions.
For example,
some terms involving the coefficients $a_{L,R,Q,U,D}$
can be eliminated by position-dependent field-phase redefinitions,
as described in Ref.\ \cite{cksm}.
Another example is provided by terms involving the coefficients 
$H_{L,U,D}$,
some of which can absorb through field-normalization redefinitions
certain other terms involving the coefficients 
$c_{L,R,Q,U,D}$.
These examples have specific interesting implications
for the quantum-electrodynamics limit 
of the standard-model extension,
and their explicit forms for that case 
are given in section III below.
Useful nonlinear field redefinitions
might also exist in principle,
but these are typically more difficult to implement meaningfully
because they may represent (noncanonical) transformations 
between different physical systems 
rather than reinterpretations of the same physics.

We next consider the issue of electroweak 
SU(2) $\times$ U(1) symmetry breaking.
The static potential for the gauge and Higgs fields 
can be extracted from the lagrangian terms given above
for the standard-model extension.
It is possible to work in unitary gauge as usual,
since the Lorentz-breaking terms do not affect the 
gauge structure of the theory.
The analysis is somewhat more complicated 
than the conventional case,
as it involves additional terms 
depending on the coupling coefficients
$k_{\ph\ph}$, $k_{\ph W}$, $k_\ph$, $k_W$, $k_2$, and $k_0$.
In principle,
there are also contributions from the SU(3) sector,
but these decouple from the Higgs field
and so the gluon expectation values 
can be taken to be zero as usual.
As mentioned above,
the terms $k_2$ and $k_0$ are expected to vanish
for consistency of the minimal theory,
and so we assume this in what follows.
In fact,
a nonzero $k_2$ would have no effect 
on the vacuum values of the fields,
but the linear instability 
that would be introduced by a nonzero $k_0$ 
would exclude a stable vacuum 
in the absence of other (nonlinear) effects.

Extremizing the static potential produces 
five simultaneous equations.
Three of these are satisfied if the expectation values
of $W_\mu^\pm$ and the photon $A_\mu$ vanish.
The other two equations can be solved algebraically
for the expectation values of the Higgs and $Z^0_\mu$ fields.
In the general case,
both of these are nonzero
and are given by
\beq
\vev{Z_\mu^0} = \fr 1 q {\sin2\th_W} 
(\Re \hat k_{\ph\ph})^{-1}_{\mu\nu} k^\nu_\ph
\quad ,
\label{zsol}
\eeq
\beq
\vev{r_\ph} = 
a \left( 1 - \fr 1 {\mu^2} 
(\Re \hat k_{\ph\ph})^{-1}_{\mu\nu} k^\mu_\ph k^\nu_\ph
\right)^{1/2}
\quad ,
\label{higgssol}
\eeq
where 
$\hat k_{\ph\ph}^{\mu\nu} \equiv 
\et^{\mu\nu} + k_{\ph\ph}^{\mu\nu}$
and 
$a \equiv \sqrt{6 \mu^2/\la}$.
Note that the quantity
$(\Re \hat k_{\ph\ph})^{-1}_{\mu\nu}$
always exists
when the Lorentz violation is small,
$|(k_{\ph\ph})^{\mu\nu}| \ll 1$.
Note also that $\vev{r_\ph}$ is a scalar
under both particle and observer Lorentz transformations,
so quantities such as the fermion mass parameters
remain scalars despite the presence of Lorentz breaking. 

As might be anticipated,
the above pattern of expectation values leaves unbroken
the electromagnetic U(1) symmetry, 
and it can be shown that fluctuations 
about the extremum are stable.
When substituted into the lagrangian 
for the standard-model extension,
the unconventional nonzero expectation value
for the field $Z^0_\mu$ 
generates some additional 
CPT- and Lorentz-violating contributions.
However,
these are all of the same form 
as other CPT- and Lorentz-violating terms
already present in the theory,
so they can be absorbed into existing coupling coefficients.

Some analyses of experimental tests of the standard-model extension
involving flavor-changing oscillations in neutral mesons
have been performed in Refs.\ \cite{ckpv,bexpt,ak}.
Tests at the level of quantum electrodynamics are mentioned below.
Note that some bounds on both the fermion and the gauge sectors 
might be obtained from available experimental information 
about the $Z^0_\mu$ and perhaps the $W^\pm_\mu$.
Such limits would be of interest in their own right,
although it seems likely that they would be much weaker 
than required to detect suppressed Lorentz violation 
at the levels estimated in this work.

\section{EXTENDED QUANTUM ELECTRODYNAMICS}

In much the same way that
conventional quantum electrodynamics (QED)
can be obtained from the usual standard model,
a generalized quantum electrodynamics 
incorporating Lorentz-breaking terms can be extracted 
from the standard-model extension given in section II.
This is of particular interest because
QED has been tested to high precision in a variety of experiments,
some of which may tightly constrain
the coupling coefficients of the possible Lorentz-violating terms.

A straightforward way to obtain the extended QED
is as follows.
After the SU(2) $\times$ U(1) symmetry breaking,
set to zero the fields $G_\mu$ for the gluons,
$W^\pm_\mu$, $Z^0_\mu$ for the weak bosons,
and the physical Higgs field
(but not the expectation value of the Higgs doublet,
which generates fermion masses). 
The only remaining boson is the photon,
mediating the electromagnetic interactions.
The neutrinos are charge neutral,
so they decouple and can be discarded.
The resulting theory is an extended QED 
describing the electromagnetic interactions 
of quarks and (charged) leptons.
It is expected to inherit from the standard-model extension
various attractive features mentioned in the introduction,
including 
U(1) gauge invariance,
energy-momentum conservation,
observer Lorentz invariance,
hermiticity,
microcausality,
positivity of the energy,
and power-counting renormalizability.

Denote the standard four-component lepton fields by $l_A$
and their masses by $m_A$,
where $A = 1,2,3$ corresponds to
electron, muon, tau, respectively.
Then,
the lagrangian for the conventional QED 
of leptons and photons is
\beq
\cl^{\rm QED}_{\rm lepton-photon} = 
\half i \overline{l}_A \ga^\mu \lrDmu l_A
- m_A \overline{l}_A l_A
- \frac 1 4 F_{\mu\nu}F^{\mu\nu}
\label{qedleptonphoton}
\quad .
\eeq
In this equation and throughout what follows,
$D_\mu \equiv \prt_\mu + i q A_\mu$ and
the field strength $F_{\al\be}$ is defined by
\beq
F_{\al\be} \equiv \prt_\al A_\be - \prt_\be A_\al
\quad ,
\label{Fdef}
\eeq
as usual.

The standard-model extension generates 
additional terms that violate Lorentz symmetry.
The CPT-even terms involving the lepton fields are
\bea
\cl^{\rm CPT-even}_{\rm lepton} &=& 
- \half (H_l)_{\mu\nu AB} \overline{l}_A \si^{\mu\nu} l_B
\nonumber\\ &&
+ \half i (c_l)_{\mu\nu AB} \overline{l}_A \ga^{\mu} \lrDnu l_B
\nonumber\\ &&
+ \half i (d_l)_{\mu\nu AB} \overline{l}_A \ga_5 \ga^\mu \lrDnu l_B
\label{lorviolqed}
\quad .
\eea
In this equation, 
the coupling coefficients
$(H_l)_{\mu\nu AB}$ are antisymmetric in spacetime indices
and have dimensions of mass.
They arise from the coefficients in Eq.\ \rf{loryukawa}
following gauge-symmetry breaking,
and they are hermitian in generation space.
The hermitian dimensionless couplings 
$(c_l)_{\mu\nu AB}$ and $(d_l)_{\mu\nu AB}$
could in principle have 
both symmetric and antisymmetric spacetime components
but can be taken traceless.
They arise from the expressions \rf{lorvioll}.

The CPT-odd terms involving the lepton fields are
\beq
\cl^{\rm CPT-odd}_{\rm lepton} = 
- (a_l)_{\mu AB} \overline{l}_A \ga^{\mu} l_B
- (b_l)_{\mu AB} \overline{l}_A \ga_5 \ga^{\mu} l_B
\label{cptviolqed}
\quad .
\eeq
The couplings $(a_l)_{\mu AB}$ and $(b_l)_{\mu AB}$
are hermitian and have dimensions of mass.
They arise from Eq.\ \rf{cptvioll}.
Note that
imposing individual lepton-number conservation 
in both the above equations
would make all the coupling coefficients
diagonal in flavor space.

In the pure-photon sector,
there is one CPT-even Lorentz-violating term: 
\beq
\cl^{\rm CPT-even}_{\rm photon} =
-\frac 1 4 (k_F)_{\ka\la\mu\nu} F^{\ka\la}F^{\mu\nu}
\quad .
\label{lorqed}
\eeq
The coupling 
$(k_F)_{\ka\la\mu\nu}$ 
arises from Eq.\ \rf{lorgauge}
and is real and dimensionless.
Without loss of generality
it can be taken as double traceless,
since any trace component would serve merely
to redefine the kinetic term 
and hence is just a field renormalization.
We disregard a conceivable $\th$-type term proportional to 
$F_{\ka\la} \ep^{\ka\la\mu\nu} F_{\mu\nu}$,
which might arise from a totally antisymmetric component of $k_F$,
on the grounds that it is a total derivative.
The coupling $k_F$ therefore
can be taken to have the symmetries of the Riemann tensor.

There is also a CPT-odd pure-photon term: 
\beq
\cl^{\rm CPT-odd}_{\rm photon} =
+ \half (k_{AF})^\ka \ep_{\ka\la\mu\nu} A^\la F^{\mu\nu} 
\quad ,
\label{cptqed}
\eeq
where the coupling coefficient
$(k_{AF})^\ka$
is real and has dimensions of mass.
This term arises from the CPT-odd gauge sector \rf{cptgauge}
of the standard-model extension.
As mentioned in the previous section,
it has some theoretical difficulties 
associated with negative contributions to the energy
and it therefore seems likely to be absent in practice.
It is included in what follows
so that we can discuss explicitly its difficulties
and some related issues involving radiative corrections.
Note also that the excluded destabilizing linear term 
in $B_\mu$ in the standard-model extension would,
if present,
generate a corresponding linear term $- (k_A)_\ka A^\ka$
in Eq.\ \rf{cptqed},
where $(k_A)_\ka$ is a real coupling
with dimensions of mass cubed.
Certain issues involving this term
are addressed in sections IV A and V.

The QED limit obtained from the standard-model extension
also has a quark sector.
This has the same general form as the lepton sector 
given by Eqs.\ \rf{qedleptonphoton}, \rf{lorviolqed},
and \rf{cptviolqed},
except that six quark fields replace the three leptons
and so twice as many Lorentz-violating couplings occur.
Note that the lepton and quark sectors 
are coupled only through the photon:
the gauge invariance of the standard-model extension
excludes couplings mixing leptons and quarks. 

The extended QED of leptons and photons given in 
Eqs.\ \rf{qedleptonphoton} - \rf{cptqed}
should suffice for certain applications where
the asymptotic states are leptons or photons
and the strong and weak interactions play a negligible role,
including a variety of existing or proposed 
high-precision experiments involving leptons.
Interesting options for such experiments are to establish 
the possible signals of Lorentz violation 
suggested by the extended QED 
and to place bounds on the associated coupling coefficients.
For example,
promising possibilities involving the muon 
include accurate measurements of $g-2$ 
such as those underway at the Brookhaven muon ring
\cite{bnl}
and sensitive tests for the decay $\mu\to e\ga$.
There are also a variety of other comparisons
involving heavy leptons that are potentially of interest
\cite{pdg}.
These issues lie beyond the scope of the present work
and will be addressed elsewhere.

For certain experiments,
it suffices to consider another limiting case of the theory:
the extended QED including only electrons, positrons and photons.
This limit can be extracted from the 
lagrangian terms for the extended QED of leptons and photons
by setting to zero the muon and the tau fields.
Denoting the four-component electron field by $\ps$
and the electron mass by $m_e$,
the usual QED lagrangian for electrons and photons is
\beq
\cl^{\rm QED}_{\rm electron} = 
\half i \overline{\ps} \ga^\mu \lrDmu \ps 
- m_e \overline{\ps} \ps
- \frac 1 4 F_{\mu\nu}F^{\mu\nu}
\label{qed}
\quad .
\eeq
In the Lorentz-violating sector,
the pure-photon terms are still given by Eqs.\ \rf{lorqed}
and \rf{cptqed}.
However,
the CPT-even terms in the fermion sector become 
\bea
\cl^{\rm CPT-even}_{\rm electron} &=& 
- \half H_{\mu\nu} \overline{\ps} \si^{\mu\nu} \ps 
\nonumber\\ &&
+ \half i c_{\mu\nu} \overline{\ps} \ga^{\mu} \lrDnu \ps 
\nonumber\\ &&
+ \half i d_{\mu\nu} \overline{\ps} \ga_5 \ga^\mu \lrDnu \ps
\label{lorvioleqed}
\quad ,
\eea
while the CPT-odd ones become 
\beq
\cl^{\rm CPT-odd}_{\rm electron} = 
- a_{\mu } \overline{\ps} \ga^{\mu} \ps 
- b_{\mu} \overline{\ps} \ga_5 \ga^{\mu}\ps 
\label{cptvioleqed}
\quad .
\eeq
The real coupling coefficients
$a$, $b$, $c$, $d$, and $H$
are the $(1,1)$-flavor components 
of the corresponding coefficients 
in the extended QED of leptons and photons
and inherit the corresponding dimensions 
and Lorentz-transformation properties.

In addition to the expressions
given in Eqs.\ \rf{lorqed} - \rf{cptvioleqed}
for the extended QED of electrons, positrons, and photons,
other Lorentz-violating terms can be envisaged
that are compatible with U(1) charge symmetry,
renormalizability, 
and an origin in spontaneous Lorentz breaking
but that \it cannot \rm be obtained 
as a reduction from the standard-model extension.
All such terms would be CPT odd.
They would have the form 
\bea
\cl^{\rm extra}_{\rm electron} &=& 
\half i e_{\nu} \overline{\ps} \lrDnu \ps 
- \half f_{\nu} \overline{\ps} \ga_5 \lrDnu \ps 
\nonumber\\ &&
+ \frac 1 4 i g_{\la\mu\nu} \overline{\ps} \si^{\la\mu} \lrDnu \ps
\label{extra}
\quad ,
\eea
where the couplings $e_\mu$, $f_\mu$ and $g_{\la\mu\nu}$
are real and dimensionless.
The reason such terms are absent from the 
expressions obtained above is that 
all putative renormalizable terms in the standard-model extension
that could generate Eq.\ \rf{extra}
are directly incompatible with the electroweak structure.
However,
it is possible that
nonrenormalizable higher-dimensional operators
in the effective lagrangian 
obeying SU(2) $\times$ U(1) symmetry 
and involving the Higgs field 
might generate the expressions \rf{extra}
when the Higgs field acquires its vacuum expectation value.
According to standard lore and the discussion in section II,
such operators would be expected to be highly suppressed 
relative to those we have listed for the standard-model extension.
This suppression should remain in force
at the level of the extended QED,
which means any terms of the form \rf{extra} 
would be expected to have coupling coefficients much
smaller than the other terms we consider.
Similar considerations apply to possible extra terms 
that might appear in the heavy-lepton and quark sectors
of the extended QED.

Next,
we address the issue of field redefinitions 
within the context of 
the extended QED of electrons, positrons, and photons.
We have found several cases to be especially useful.
One is a linear phase redefinition 
of the form $\ps = \exp(- i a\cdot x) \ch$,
which eliminates the term 
$- a_{\mu } \overline{\ps} \ga^{\mu} \ps$ 
from Eq.\ \rf{cptvioleqed}.
This is equivalent to shifting the zeros
of energy and momentum for electrons and positrons
\cite{cksm}.
We therefore expect no observable effects
from a nonzero $a_\mu$ in any QED experiment.

Another useful class of redefinitions
involves field renormalizations depending 
on coupling coefficients.
For a fermion field $\ps$,
consider the redefinition
\beq
\ps = (1 + v \cdot \Ga) \ch
\quad ,
\label{redef}
\eeq
where $\Ga$ is one of 
$\ga^\mu$, $\ga_5\ga^\mu$, $\si^{\mu\nu}$
and $v$ is a combination of coupling coefficients with
appropriate spacetime indices.
This set of redefinitions can be used 
to obtain several useful approximate results,
valid to first order in the (small) coupling coefficients.
One is that the combination 
$\ep^{\mu\nu\al\be}H_{\al\be} + m (d^{\mu\nu}- d^{\nu\mu})$
can be eliminated 
and hence is unobservable at leading order 
in any QED experiment.
Only the orthogonal linear combination is physical
at this level.
Another is that the antisymmetric component of $c_{\mu\nu}$ 
can be eliminated to first order.
Similarly,
even if the extra terms with coefficients $e_\mu$
and $f_\mu$ in Eq.\ \rf{extra} should appear,
they could be eliminated to first order by a 
combination of field redefinitions.
The same is true of the trace components of the extra term
with coefficient $g_{\la\mu\nu}$,
while the totally antisymmetric component of this term
can be absorbed into $b_\mu$ to first order.
Combining all these results,
it follows that at leading order
in the extended QED of electrons, positrons, and photons
the only observable coupling coefficients
can be taken as 
$b_\mu$, $H_{\mu\nu}$,
the symmetric components of $c_{\mu\nu}$ and $d_{\mu\nu}$,
and possibly the traceless mixed-symmetry components 
of the extra coefficient $g_{\la\mu\nu}$.

So far in this section 
we have considered various forms of 
extended QED that emerge 
as limits of the standard-model extension.
For some purposes,
it can be useful to work within an effective extended QED
valid for a free fermion
that is a composite of leptons and quarks,
such as a nucleon, atom, or ion.
For a single fermion field $\ps$ of this type,
the effective lagrangian would then have the same form
as that of the extended QED for electrons, positrons, and photons.
A description of this type is useful for investigations
of the implications of high-precision experiments
on composite fermions,
such as comparative tests of proton properties
or searches for a neutron electric-dipole moment. 
In principle,
extra terms of the form \rf{extra} could appear as 
a result of the interactions among the fermion constituents,
but in the effective theory the coupling coefficients
of such terms would involve combinations 
of the constituent coupling coefficients
with the interaction coupling constants
and might therefore be expected to be absent at leading order
in many cases.

Some possible experimental signals from extended QED
are investigated in Ref.\ \cite{bkr}.
Certain high-precision tests
that could be performed with present technology are considered,
and the attainable bounds on Lorentz-breaking 
coupling coefficients are estimated.
The tests involve comparative measurements
of anomalous magnetic moments and charge-to-mass ratios
for particles and antiparticles confined in a Penning trap
\cite{pennexpts}.
They typically have the potential to bound the coupling coefficients
of Lorentz- and CPT-violating terms 
at a level close to that expected from Planck-scale suppression.
For example,
the spacelike components of the coefficient $b_\mu$
control the appropriate figure of merit
for experiments comparing the anomalous magnetic moments
of the electron and positron.
This figure of merit
can be bounded to about one part in $10^{20}$,
which is comparable to the ratio of $m_e/M$ of the
electron mass to the Planck scale.

\section{THE PURE-PHOTON SECTOR}

In this section,
we focus on the pure-photon sector
of the extended QED.
We examine some theoretical implications 
of the existence of Lorentz- and CPT-violating terms
and address some experimental issues.

\subsection{Lagrangian and Energy-Momentum Tensor}

The lagrangian of interest,
which is U(1) gauge invariant by construction,
is a combination of the photon term in 
Eq.\ \rf{qed}
with the expressions \rf{lorqed} and \rf{cptqed}.
It is 
\bea
\cl^{\rm total}_{\rm photon} &=&
- \frac 1 4 F_{\mu\nu}F^{\mu\nu}
-\frac 1 4 (k_F)_{\ka\la\mu\nu} F^{\ka\la}F^{\mu\nu}
\nonumber\\ &&
+ \half (k_{AF})^\ka \ep_{\ka\la\mu\nu} A^\la F^{\mu\nu} 
\quad .
\label{total}
\eea
Some properties of the coupling coefficients $k_F$ and $k_{AF}$
are described following Eqs.\ \rf{lorqed} and \rf{cptqed}.
For certain calculations,
it is useful to decompose the coefficient $k_F$ into
its two Lorentz-irreducible pieces,
one with 10 independent components
analogous to the Weyl tensor in general relativity
and one with 9 components 
analogous to the trace-free Ricci tensor.
Only one of the 19 total independent components of $k_F$
(the 00 component of the trace-free Ricci-tensor analogue)
and one of the four independent components 
of the coefficient $k_{AF}$ 
(the timelike component $k_{AF}^0$)
are associated with terms invariant under (particle) rotations.

Some insight into the structure of the lagrangian
can be obtained by expressing it in terms
of the potentials $\ph$, $\vec A$ 
and the fields $\vec E$, $\vec B$.
We find
\bea
\cl^{\rm total}_{\rm photon} &=& 
\half (\vec E^2 - \vec B^2)
+\half \al (\vec E^2 + \vec B^2)
\nonumber \\ && 
+ \half \be_E^{jk} E^jE^k
+ \half \be_B^{jk} B^jB^k
+ \half \be_{EB}^{jk} E^jB^k
\nonumber \\
&& +k_{AF}^0 \vec A \cdot \vec B 
- \ph \vec k_{AF} \cdot \vec B
\nonumber \\ && 
+ \vec k_{AF} \cdot (\vec A \times \vec E)
\quad .
\label{totaleb}
\eea
Here and throughout this work, 
$j,k,\ldots = 1,2,3$ are spatial indices.
The real coefficients $\al$ and 
$\be_{E}^{jk}$, $\be_{B}^{jk}$, $\be_{EB}^{jk}$
are various combinations of the couplings
$(k_F)_{\ka\la\mu\nu}$ appearing in Eq.\ \rf{total}.
Disregarding as before any total-derivative effects,
the $\be_{E}^{jk}$, $\be_{B}^{jk}$, $\be_{EB}^{jk}$
are traceless.
Note that all possible quadratic combinations of 
the electric and magnetic fields appear.
Only two terms,
involving $\al$ and $k_{AF}^0$,
preserve (particle) rotational invariance.
Note also that a rescaling without physical consequences
can be performed to obtain 
a standard normalization of the electric field $\vec E$.
This produces a lagrangian of the same general form 
as Eq.\ \rf{totaleb}
except that the Lorentz-breaking term proportional to
$(\vec E^2 + \vec B^2)$
is replaced with one proportional to $\vec B^2$ alone. 

The canonical energy-momentum tensor 
can be constructed following the standard procedure.
This tensor can be partially symmetrized,
but complete symmetrization is impossible 
because there is an antisymmetric component 
that cannot be written as a total derivative.
A relatively elegant expression can be obtained
by adding judiciously chosen total-derivative terms,
which leave unchanged the physics.
Denoting the resulting energy-momentum tensor by
$\Th^{\mu\nu}$,
we find
\bea
\Th^{\mu\nu} &=&
-F^{\mu\ga}F^\nu_{\pt{\nu}\ga} 
+ \frac 1 4 \et^{\mu\nu} F_{\al\be}F^{\al\be}
\nonumber \\ &&
-(k_F)^{\al\be\mu\ga}F^\nu_{\pt{\nu}\ga} F_{\al\be}
+ \frac 1 4 \et^{\mu\nu} (k_F)_{\al\be\ga\de}F^{\al\be} F^{\ga\de}
\nonumber \\ &&
+(k_{AF})^\nu A_\al \widetilde F^{\al\mu}
\quad .
\label{enmom}
\eea
Here, 
we define
\beq
\widetilde F^{\mu\nu} = \ep^{\ka\la\mu\nu}F_{\ka\la}/2
\quad 
\label{Ftilde}
\eeq
to be the dual field strength.

The energy-momentum tensor obeys 
the usual conservation relation,
\beq
\prt_\mu \Th^{\mu\nu} = 0
\quad .
\label{enmomcons}
\eeq
In addition to the gauge-invariant and symmetric contributions
to $\Th^{\mu\nu}$,
which include the conventional pieces among others,
there are additional terms involving the coefficient $k_F$ 
that are gauge invariant but asymmetric.
The term with $k_{AF}$ is neither gauge invariant nor symmetric.
Under a gauge transformation,
an additional total-derivative term appears.
The presence of an antisymmetric component in $\Th^{\mu\nu}$ 
implies that care is required in physical interpretations
of the energy-momentum behavior.
Although $\Th^{j0}$ can be regarded
as the components of a generalized Poynting vector,
its volume integral is no longer conserved
and cannot be identified with the conserved 
volume integral of the components $\Th^{0j}$
of the momentum density.
These features are a direct consequence
of the presence of the background expectation values
of tensor fields,
represented in the low-energy theory by the coupling coefficients
$k_F$ and $k_{AF}$.

The energy density is given by the component $\Th^{00}$.
Inspection shows it can be written in the form:
\bea 
\Th^{00} &=& 
\half (\vec E^2 + \vec B^2)
\nonumber\\ &&
- (k_F)^{0j0k} E^j E^k
+ \frac 1 4 (k_F)^{jklm} \ep^{jkp} \ep^{lmq} B^p B^q
\nonumber\\ &&
- (k_{AF})^0 \vec A \cdot \vec B
\quad .
\label{endens}
\eea
If $k_{AF}$ vanishes and $k_F$ is small,
$\Th^{00}$ is nonnegative.
This can be seen as follows.
The combination of the usual energy density 
with the terms proportional to $k_F$
can be viewed as a bilinear form $x^T M x$
generated from a matrix $M$
in the six-dimensional space $x^T \equiv (\vec E, \vec B)$.
The matrix $M$ is symmetric and $3\times 3$-block diagonal,
since no cross terms in $\vec E$ and $\vec B$ 
appear in $\Th^{00}$.
Observer rotation invariance can be used to diagonalize
the upper $3\times 3$ block
associated with the electric field.
Since $k_F$ is small,
the three diagonal entries are of the form $\half - O(k_F) >0$,
so the contribution to $\Th^{00}$ 
from the electric field is nonnegative in any frame.
A similar argument shows that
the contribution from the lower $3\times 3$ block,
associated with the magnetic field,
is also nonnegative.
The conserved energy $\cE$
of a field configuration,
obtained by integrating $\Th^{00}$ over all space,
is therefore also nonnegative.

If instead $k_F$ vanishes and $k_{AF}$ is small, 
the contribution to $\cE$ can be written in the form
\bea
\cE & \equiv & 
\int d^3 x ~\Th^{00}
\nonumber \\  
&=& \half \int d^3 x \left(
\vec E^2 + [ \vec B - (k_{AF})^0 \vec A ]^2  
\right.  
\nonumber \\ && 
\left.  
\qquad\qquad\qquad\qquad 
- [(k_{AF})^0]^2 \vec A^2 \right)
\quad .
\label{cE}
\eea
The last term is nonpositive and so can in principle 
introduce an instability in the theory
\cite{cfj}.
Note that a similar situation would hold for
the linear term 
$- (k_A)_\ka A^\ka$ in the lagrangian
that was discarded in section III,
for which the energy density 
$(k_A)_0 \ph - \vec k_A \cdot \vec A$
could also be negative.
The appearance of negative contributions to the energy
is unsatisfactory from a theoretical viewpoint.

It might seem tempting to resolve this issue
by requiring that only the spacelike components of $k_{AF}$ 
are nonzero,
so that the terms involving $(k_{AF})^0$ are absent from
Eqs.\ \rf{endens} and \rf{cE}.
However,
this condition depends on the observer frame,
so even an infinitesimal boost to another observer frame
would reintroduce the instability.
A somewhat more interesting option might be 
to combine the vanishing of $(k_{AF})^0$ 
with the introduction of a (small) photon mass,
perhaps arising from a hitherto unobserved spontaneous
breaking of the electromagnetic U(1) gauge symmetry.
This would eliminate the linear instability
and in principle might also produce a contribution canceling
the negative term appearing in Eq.\ \rf{cE},
although perhaps only for a physically reasonable range 
of observer boosts determined by the size of the photon mass
and the magnitude of the components of $\vec k_{AF}$.
Although some form of this idea 
might be made physically acceptable,
we are restricting ourselves here
to minimal modifications of the usual standard model
and so we disregard this possibility in the present work. 

In the absence of a complete demonstration 
of a consistent alternative interpretation, 
one option might be to discard the term \rf{cptqed} 
depending on $k_{AF}$.
This is possible at the classical level,
but at the quantum level one might 
expect radiative corrections to induce it.
We return to this question in section V,
meanwhile keeping the term \rf{cptqed} 
in the analysis for completeness.

\subsection{Solution of Equations of Motion}

The equations of motion arising from the lagrangian \rf{total}
are:
\beq
\prt_\al F_\mu^{\pt{\mu}\al} 
+ (k_F)_{\mu\al\be\ga}\prt^\al F^{\be\ga}
+ (k_{AF})^\al \ep_{\mu\al\be\ga} F^{\be\ga} 
= 0
\quad .
\label{eqmot}
\eeq
These equations are the 
Lorentz-breaking extensions of the usual
inhomogeneous Maxwell equations
in the absence of sources,
$\prt_\mu F^{\mu\nu} = 0$.
By virtue of its conventional definition in Eq.\ \rf{Fdef},
the field strength $F^{\mu\nu}$ satisfies
the usual homogeneous Maxwell equations
\beq
\prt_\mu {\widetilde F}^{\mu\nu} = 0
\quad ,
\label{homeqmot}
\eeq
where ${\widetilde F}^{\mu\nu}$ is given in Eq.\ \rf{Ftilde}.

An important feature
of the equations \rf{eqmot} and \rf{homeqmot}
is their linearity in $F_{\mu\nu}$ and hence in $A_\mu$.
The Lorentz-violating terms thereby avoid
the complications of nonlinear modifications
to the Maxwell equations,
which are known to occur in some physical situations
such as nonlinear optics or 
when vacuum-polarization effects are included.
Another feature is that
the extra Lorentz-violating terms involve both the
electric and the magnetic fields,
as well as their derivatives.
As a result,
the equations \rf{eqmot} bear some resemblance
to the usual Maxwell equations in moving media,
for which the boost 
causes the electric and magnetic fields to mix.
Note that the coefficients determining this mixing
are directly dependent on the velocity of the medium
and so change with the inertial frame.
Similarly,
for the Lorentz-violating case of interest here,
a change of observer frame changes the coupling coefficients.
Some other useful analogies between the equations \rf{eqmot}
and those of conventional electrodynamics in macroscopic media
are described in section IV C.

The equations of motion \rf{eqmot} and \rf{homeqmot}
depend only on $F_{\mu\nu}$ and so,
as expected,
they are gauge invariant under the standard 
U(1) gauge transformations
\beq
A_\mu \to A_\mu - \fr 1 q \prt_\mu \La
\quad .
\label{gauge}
\eeq
As in conventional electrodynamics,
the presence of gauge symmetry affects 
the interpretation and solution of the equations of motion.
We first consider a treatment in terms of the potentials $A_\mu$
and then one for the field strengths $\vec E$, $\vec B$.

Taking the potentials $A_\mu$ as basic,
the four equations \rf{homeqmot} are directly satisfied.
This appears to leave four equations \rf{eqmot} 
for four unknowns $A_\mu$.
However,
just as in conventional Maxwell electrodynamics,
the conjugate momentum to $A_0$ vanishes
because $\prt_0 A^0$ is missing from the lagrangian \rf{total},
so the theory has a Dirac primary constraint.
In the conventional case
it then follows from the identity 
$\prt_\nu\prt_\mu F^{\mu\nu}\equiv 0$,
which is associated with current conservation 
when sources are present,
that the equation of motion associated with $A_0$
plays the role of an initial condition.
The same conclusion holds here
because when acted on by $\prt_\mu$
the left-hand side of Eq.\ \rf{eqmot}
also vanishes identically.
This leaves three equations of motion and a constraint
for four variables.
One combination of variables can be fixed by a gauge choice.
The constraint 
then leaves two independent degrees of freedom.

Despite the parallels with conventional electrodynamics,
the gauge-fixing process involves some interesting differences.
For example,
there is normally an equivalence between
the Coulomb gauge 
$\vec\nabla \cdot \vec A = 0$,
the temporal gauge 
$A^0 = 0$,
and one of the members of the family of Lorentz gauges
$\prt_\mu A^\mu = 0$.
When Lorentz-violating effects are included,
these three gauge choices become inequivalent.
For example,
$A^0$ typically is nonzero
if the Coulomb gauge
$\vec\nabla \cdot \vec A = 0$ is imposed.

More insight about the wave motion implied by Eq.\ \rf{eqmot} 
can be gained with the ansatz
\beq
A_\mu (x) \equiv 
\ep_\mu (p) \exp(-ip_\al x^\al) 
\quad ,
\label{momspaceA}
\eeq
where $p^\mu \equiv (p^0, \vec p)$
can be regarded as the frequency and wave vector of the mode
or as the associated energy and momentum
(which can be distinct from the conserved energy and momentum
obtained from the energy-momentum tensor).
Note that taking the real part  
is understood, as usual.
The equations of motion \rf{eqmot} generate
the momentum-space equation 
\beq
M^{\al\de}(p) A_\de = 0
\quad ,
\label{momeqmot}
\eeq
where the matrix $M^{\al\de}(p)$ is
\bea
M^{\al\de}(p) &\equiv &
\et^{\al\de} p^2 - p^\al p^\de
- 2 (k_F)^{\al\be\ga\de}p_\be p_\ga
\nonumber\\ &&
- 2 i (k_{AF})_\be \ep^{\al\be\ga\de} p_\ga 
\quad .
\label{Mmatrix}
\eea
This $4\times 4$ complex-valued matrix
is hermitian because the first three terms
are real and symmetric while the last is imaginary and antisymmetric.
Its determinant can be shown to vanish 
identically for all $p^\mu$,
a feature related to the gauge freedom.
The conventional result
is recovered when the coefficients $k_F$ and $k_{AF}$ vanish.

Once a gauge choice is imposed,
the relation \rf{momeqmot} provides  
a set of complex-valued equations for $A_\de$.
The differences between various gauge choices 
that normally are equivalent can be seen explicitly at this stage.
For example,
$M^{\al\de}(p)A_\de$ is not proportional to $A^\al$ in
the Lorentz gauges, 
the Coulomb gauge leaves a nontrival equation for $A^0$,
and the temporal gauge generates an involved constraint 
on $\vec\nabla \cdot \vec A$.
In practice,
Eq.\ \rf{momeqmot} then reduces to a (sub)set of equations 
involving an effective matrix $M_{\rm eff}(p)$
with explicit form dependent on the gauge choice.
The requirement for existence of nonzero solutions 
can be obtained from a condition of the type 
$\det M_{\rm eff}(p) = 0$.
With fixed coefficients $k_F$ and $k_{AF}$,
this condition then determines $p^0$ as a function of $\vec p$. 
Since $M^{\mu\al}(p)$ is a $4\times 4$ matrix with entries 
quadratic in $p^\mu$,
a determinant of this type 
can produce an eighth-order polynomial in $p^0$.

In the conventional case in the Lorentz gauge, 
the polynomial reduces to one with two quadruply degenerate roots,
$p^0 = \pm |\vec p|$.
The apparent doubling of the roots 
relative to the number of variables 
can be understood by the observation that in this case 
$M^{\mu\al}(p)$ is symmetric under $p^\mu \to - p^\mu$,
so for each solution $p^0(\vec p)$ there is another
solution $-p^0(-\vec p)$.
These two solutions can be shown to be physically equivalent
by examining the real part of $A_\mu$ in Eq.\ \rf{momspaceA}.

In contrast,
in the general extended electrodynamics
the polynomial determining $p^0$ may have eight distinct roots.
Each of these could in principle 
produce a nontrivial solution for $A_\de$,
double the expected number.
In this case,
$M^{\mu\al}(p)$ is symmetric under 
the simultaneous operations
$p^\mu \to - p^\mu$ and $k_{AF} \to - k_{AF}$
(leaving $k_F$ unchanged).
Thus,
for each solution $p^0(\vec p, k_F, k_{AF})$ there is another
solution $-p^0(-\vec p, k_F, -k_{AF})$.
The sign change for the coefficient $k_{AF}$
might appear to preclude the demonstration 
of the physical equivalence of these solutions.
However,
hermiticity of $M^{\mu\al}(p)$
implies its determinant is equivalent to 
the determinant of its complex conjugate.
Since $k_{AF}$ appears only in the imaginary part of
$M^{\mu\al}(p)$,
it follows that for each solution $-p^0(-\vec p, k_F, -k_{AF})$
there is also a solution $-p^0(-\vec p, k_F, +k_{AF})$.
This is physically equivalent to the solution
$p^0(\vec p, k_F, k_{AF})$,
as before.
Thus,
the number of independent roots
is the same as the number of variables as expected,
despite the apparent complexity of the polynomial.

In the general extended electrodynamics,
neither the Coulomb gauge nor the Lorentz gauges 
significantly simplify the primary constraint.
In contrast,
the temporal gauge $A^0=0$ immediately removes one 
degree of freedom from $A_\mu$.
This gauge can be imposed by choosing 
the function $\La$ in Eq.\ \rf{gauge}
as $q\La (t, \vec x)= \int^t A^0(t^\prime, \vec x) dt^\prime$.
Note that this choice breaks observer boost invariance
but leaves unaffected the observer rotation invariance.
It reduces the primary constraint to the form
\beq
M^{0j} A^j = 0
\quad ,
\label{MA}
\eeq
where $M^{0j}$ are components in the temporal gauge
of the matrix $M^{\al\de}$ given in Eq.\ \rf{Mmatrix}.

At this stage an explicit solution could be found.
For example,
one could use two of the degrees of freedom 
of observer rotation invariance
to select a convenient coordinate system,
such as one in which $p^j \equiv (0,0,p)$.
Solving for $A^3$ from the primary constraint \rf{MA}
and substituting into the remaining three equations
of motion in \rf{momeqmot}
would then produce an identity 
and two simultaneous linear equations for $A^1$ and $A^2$.
A nontrivial solution of this pair of equations could be found 
by requiring the determinant of the system to vanish,
which in turn would generate a relation between $p^0$ and $p$.
Solving this relation must give 
two independent dispersion relations,
one for each of the two physical degrees of freedom.
The full dispersion relations in an arbitrary coordinate system
could then in principle be recovered 
by using arguments based on rotational covariance.

Rather than pursuing this approach,
we return to the 
eight equations of motion \rf{eqmot} and \rf{homeqmot}
and reconsider their treatment
taking as independent variables 
the six electric and magnetic fields.
Here,
we are interested in the properties of electromagnetic radiation,
so we work with the standard ansatz
\beq
F_{\mu\nu} (x) \equiv 
F_{\mu\nu} (p) \exp(-ip_\al x^\al) 
\quad ,
\label{momspaceF}
\eeq
where $p^\mu \equiv (p^0, \vec p)$.

The equations \rf{homeqmot},
which include the usual Faraday law and the condition 
ensuring the absence of magnetic monopoles,
are unaffected by the Lorentz breaking
and for radiation reduce as usual to
\beq
p^0 \vec B =\vec p \times \vec E 
\quad , \quad
\vec p \cdot \vec B = 0
\quad .
\label{fara}
\eeq
The first of these 
can be regarded as defining the magnetic field
once the electric field is known.
The second of these equations follows from the first,
and shows that the magnetic field remains 
transverse to $\vec p$ despite the Lorentz violation. 

The equations \rf{eqmot}
generate modified Coulomb and Amp\`ere laws
that are to be solved for $\vec E$.
A relatively straightforward procedure 
is to substitute for $\vec B$ from Eq.\ \rf{fara}.
Using the Amp\`ere law,
we thereby obtain the vector equation 
\beq
M^{jk}E^k = 0
\quad ,
\label{amp}
\eeq
where the 3$\times$3 matrix $M^{jk}$ is identical
to the $(jk)$-component submatrix of 
the matrix $M^{\al\de}$ in Eq.\ \rf{Mmatrix}.
The modified Coulomb law
can be obtained from the modified Amp\`ere law
by taking the scalar product with $\vec p$:
$p^jM^{jk}E^k = 0$.
Note that this derivation provides some insight 
about the temporal gauge $A^0 = 0$
in a treatment using $A_\mu$.
Thus, 
there is a close parallel between the two 
because $\vec E = i p^0 \vec A$ in this gauge.

To obtain an explicit solution,
it is helpful to take advantage of the
observer rotation invariance to select a coordinate system
in which key expressions are simplified.
For example,
a useful frame is the one with 
$\vec p = (0,0,p)$.
In it,
the modified Coulomb equation can be solved for $E^3$
in terms of $E^1$ and $E^2$.
Substitution of this solution into 
the three component equations \rf{amp}
produces one identity and
two simultaneous linear equations for 
$E^1$ and $E^2$.
The matrix of this system of equations is hermitian,
and the condition for a nontrivial solution
is that its determinant vanishes.
The determinant turns out to be a fourth-order polynomial
for $p^0$ as a function of $p$.
For reasons similar to those given for the $A_\mu$ case,
there are only two physically distinct solutions
of this polynomial,
one for each of the two independent degrees of freedom.

An explicit solution of the determinant condition
in the general case is involved 
because the fourth-order polynomial is
homogeneous in the small coupling coefficients.
Since $k_F$ is dimensionless
while $k_{AF}$ has dimensions of mass,
to first order in the coupling coefficients
and for $p^0\approx p \gg |k_{AF}|$ and $1\gg |k_F|$
the solution for $p^0$ as a function of $p$
must take the form of 
the sum of $p$ with a function of the 
quantities $k_F p$ and $k_{AF}$.
Indeed,
we find 
\beq
p^0 = (1 + \rh) p \pm \sqrt{(\si^2 p^2 + \ta^2)}
\quad ,
\label{disp}
\eeq
where $\rh$ and $\si^2$ are functions of the components of $k_F$
and $\ta^2$ is a function of the components of $k_{AF}$,
given by
\bea
\rh &=& 
\half
\left[
(k_F)^{0101} + (k_F)^{0202} + (k_F)^{1313}
\right.
\nonumber \\ &&
\left.
+ (k_F)^{2323} + 2(k_F)^{0113} + 2(k_F)^{0223}
\right]
\quad ,
\nonumber\\ 
\si^2 &=& 
\frac 1 4
\left[
(k_F)^{0101} - (k_F)^{0202} + (k_F)^{1313}
\right.
\nonumber \\ 
&&
\left.
- (k_F)^{2323} + 2(k_F)^{0113} - 2(k_F)^{0223} 
\right]^2
\nonumber \\ 
&&
+ \left[
(k_F)^{0102} - (k_F)^{0123} + (k_F)^{0213} + (k_F)^{1323}
\right]^2
\quad ,
\nonumber \\ 
\ta^2 &=& 
\left[
(k_{AF})^3 - (k_{AF})^0
\right]^2
\quad .
\label{abc}
\eea
The solution \rf{disp}
entangles the components of $k_F$ and $k_{AF}$ 
in a way that cannot be separated
without additional information about their relative sizes.
Note that it reduces correctly to the result of Ref.\ \cite{cfj}
in the case $k_F = 0$.

The corresponding general solutions for the vectors 
of the electric and magnetic fields are involved 
and provide little insight for present purposes,
so we omit them here.
They exhibit two physical linear polarization
vectors for $\vec E$,
each obeying a different dispersion relation.
This produces birefringence,
among other effects.
Note in particular that,
contrary to widespread assumption in the literature,
no circularly polarized solution
to the equations of motion typically exists.
An electromagnetic wave prepared 
in a state of circular polarization
would propagate as two linearly polarized components 
with distinct dispersion relations,
so an initial circularly polarized configuration
would gradually become elliptical.
These and some other interesting results 
about the wave propagation 
are discussed further in subsections IV C and IV D below.

In the remainder of this subsection,
we present a sample analytical solution
to the equations of motion for a special case
that provides further insight.
We consider the lagrangian \rf{total} with $(k_{AF})^\mu = 0$
and with the only nonzero components of $k_F$
chosen to be $(k_F)_{0j0k} = - \half \be_j\be_k$,
where the $\be^j$ are three (small) 
real dimensionless quantities,
and components related to these by the symmetries of $k_F$.
In terms of the lagrangian \rf{totaleb}
only the term involving $\be_E^{jk}$ is nonzero,
and it has a direct-product structure:
$\be_E^{jk} \equiv + \be^j \be^k$.
The lagrangian \rf{totaleb} therefore becomes
\beq
\cl^{\rm special}_{\rm photon} 
= \half ( \vec{E}^2 - \vec{B}^2) 
+ \half (\vec\be \cdot \vec E)^2
\quad .
\label{lsamp}
\eeq
This example involves only CPT-even Lorentz violation. 

The lagrangian \rf{lsamp} generates modified  
inhomogeneous Maxwell equations in the absence of sources:
\bea
\vec{\nabla} \cdot \vec{E} 
& = & - \vec{\be} \cdot \vec{\nabla}
(\vec{\be} \cdot \vec{E})
\quad ,
\nonumber \\
\vec{\nabla} \times \vec{B} - \prt_0 \vec{E} 
& = & \vec{\be}~ \prt_0 (\vec{\be} \cdot \vec{E})
\quad .
\label{msamp}
\eea
In terms of the potentials $A_\mu$ of Eq.\ \rf{momspaceA},
appropriate for describing radiation in momentum space,
these are equivalent to the vector equation 
\bea
&& 
p^0[\vec p + (\vec p \cdot \vec\be)\vec \be ] A^0
- (p^0)^2[\vec A + (\vec A \cdot \vec\be)\vec \be ]
\nonumber\\ && 
\qquad \qquad\qquad \qquad\qquad
+[\vec p^{~2} \vec A - (\vec p \cdot \vec A ) \vec p ~]
=0
\quad 
\label{asamp}
\eea
and its scalar product with $\vec p$. 

For definiteness
we proceed in Lorentz gauge,
where $A^0 = \vec p \cdot \vec A/p^0$.
According to the discussions above,
in the presence of Lorentz violation
this gauge may require nonzero $A^0$ 
and $\vec p \cdot \vec A$.
For a nontrivial solution to Eq.\ \rf{asamp},
we find two possible dispersion relations:
\bea
(p_o)^2 & = & 0 
\quad ,
\nonumber \\
(p_e)^2 & = & - \fr { (\vec{\be}\times \vec p_e)^2}
{1 + \vec{\be}^2}
\quad .
\label{dispsamp}
\eea
The first corresponds to 
an `ordinary' mode with four-momentum $p_o$
obeying the conventional dispersion relation,
while the second is an `extraordinary' mode
with four-momentum $p_e$ and a modified dispersion relation.

For a wave vector aligned along $\vec\be$,
both modes reduce to the conventional case
and exhibit normal behavior.
However,
for other alignments the properties of the two modes differ. 
For simplicity,
we restrict attention here to
the situation with wave vector orthogonal to $\vec\be$,
so $\vec p \cdot \vec \be = 0$.
In this case,
the ordinary mode $A_o^\mu$ can be chosen to satisfy 
$A_o^0 = 0$
with $\vec A_o$ parallel to $\vec p \times \vec \be$,
while the extraordinary mode must satisfy $A_e^0 = 0$
and has $\vec A_e$ aligned along $\vec \be$.
These two modes propagate with different velocities.
For example,
their group velocities $\vec v_g \equiv \vec \nabla_p p^0$ are
\beq
\vec v_{g,o} = \hat p 
\quad , \quad
\vec v_{g,e} = 
\fr 1 {\sqrt{1 + \vec{\be}^2}} ~ \hat p 
\quad .
\label{vsamp}
\eeq
For each mode,
the group and phase velocities are equal.

One consequence of the difference between 
the two modes is birefringence.
For example,
a plane-polarized monochromatic wave of frequency $p^0$
that is initially a general combination of the two modes
eventually becomes elliptically polarized.
For the electric field,
we find
\bea
\vec{E}(t, \vec x) &=& 
-p^0 
\left( c_o \hat{A}_o \sin[p^0 (r - t)]
\pt{\sqrt{1 + \vec{\be}^2}~ r}
\right.
\nonumber\\ &&
\qquad
\left.
+ c_e \hat{A}_e \sin [p^0 (\sqrt{1 + \vec{\be}^2}~ r - t)]
\right)
\quad ,
\label{Esamp}
\eea
where $r = | \vec x|$,
$\hat A_o$ is parallel to $\vec p\times\vec\be$ 
and $\hat A_e$ is parallel to $\vec\be$ as before, 
and the weights $c_o$ and $c_e$ are determined by the 
initial polarization condition. 
This shows that the presence of $\vec\be$ 
causes the wave to become elliptically polarized 
after it has traveled a distance 
\beq
r \simeq \fr {\pi} 
{2 \left( \sqrt{1 + \vec{\be}^2} - 1 \right) p^0}
\quad .
\label{rsamp}
\eeq
The magnetic field exhibits similar behavior.

The explicit expressions for the electric and magnetic fields
can be used to derive the energy density 
$\Th_e^{00}$
and the Poynting vector 
$\Th_e^{j0}$
for the extraordinary mode.
We find
\beq
\Th_e^{00} = \vec p^{~2} c_e^2 \sin^2 p_\mu x^\mu
\quad , \quad
\Th_e^{j0} = p^0 p^j c_e^2 \sin^2 p_\mu x^\mu
\quad .
\label{emsamp}
\eeq
This shows that in the present case
the velocity of energy transport 
$v_e^j \equiv \Th_e^{j0}/ \Th_e^{00}$
is identical to the group and phase velocities,
Eq.\ \rf{vsamp}.
Some comments about the various velocities
in the general case are made in the next subsection.

\subsection{Analogy to Macroscopic Media}

In subsection IV B,
an approximate analogy was noted between the
equations of motion for 
the Lorentz-breaking extension of electrodynamics
and those for electrodynamics in moving media.
In this section,
we introduce some useful quantitative analogies 
between the extended electrodynamics
and the electrodynamics of macroscopic media.
These can be used to gain further insight 
about the nature of the extended electrodynamics
with Lorentz breaking.

Consider first the situation in position space,
where the relevant equations are \rf{eqmot} and \rf{homeqmot}.
We have already noted that Eqs.\ \rf{homeqmot}
take the same form as in conventional electrodynamics. 
The idea is to define new quantities $\vec D$ and $\vec H$ 
such that the forms of Eqs.\ \rf{eqmot}
becomes identical to those of the Maxwell equations
in material media.
It turns out that it suffices to introduce 
an effective displacement current $\vec D$
and an effective magnetic field $\vec H$
having linear dependence on the electric field $\vec E$,
the magnetic induction $\vec B$,
the vector potential $\vec A$,
and the scalar potential $A^0$.

We find that the definitions
\bea
D^j &=& E^j
- 2 (k_F)^{0j0k}E^k
+ (k_F)^{0jkl} \ep^{klm} B^m
\nonumber \\ 
&& \qquad\qquad
+2 \ep^{jkl} (k_{AF})^k A^l
\quad ,
\nonumber \\ 
H^j &=& B^j
+ \half (k_F)^{pqrs} \ep^{pqj} \ep^{rsk} B^k
- (k_F)^{0mkl} \ep^{jkl} E^m
\nonumber \\ 
&& \qquad\qquad
-2 (k_{AF})^0 A^j
+2 (k_{AF})^j A^0
\quad 
\label{DHdef}
\eea
reproduce the usual Maxwell equations in material media.
The analogy can therefore be used to gain insight
into those properties of the extended electrodynamics
that are directly associated with the equations of motion.
However,
caution is required in applying other concepts 
of conventional electrodynamics.
For example,
it turns out that if $k_{AF} \neq 0$
then the conventional expressions for the energy density and
Poynting vector in terms of $\vec D$ and $\vec H$
fail to reproduce completely the true energy density 
$\Th^{00}$ and Poynting vector $\Th^{j0}$ 
in the extended theory.
If $k_{AF} = 0$, 
in contrast,
the correct expressions are indeed reproduced 
by the analogy.

The above analogy is useful for general discussions
of the properties of the extended electrodynamics.
However,
it becomes somewhat cumbersome
for certain considerations involving radiation.
We have developed a second analogy
that is of more direct use 
when the fields are converted to momentum space
through Eq.\ \rf{momspaceF}.
It turns out that 
the equations of motion can then be correctly reproduced
by defining an effective displacement current $\vec D (p)$
through
\beq
D^j = \ep^{jk} E^k
\quad ,
\label{Ddef}
\eeq
where $\ep^{jk}$ is a hermitian effective permittivity
given by
\beq
\ep^{jk} \equiv
\de^{jk}
+ \fr 2 {(p^0)^2}
(k_F)^{j\be\ga k}p_\be p_\ga
+ \fr {2i} {(p^0)^2}
(k_{AF})_\be \ep^{j\be\ga k} p_\ga .
\label{perm}
\eeq
In particular,
it is unnecessary to introduce 
an effective magnetic field $\vec H$
distinct from $\vec B$.
This second analogy is therefore different from the first.
Note that again the correct energy density and Poynting vectors
cannot be obtained directly by substitution into
the conventional formulae.
Nonetheless,
the analogy is valuable because 
it permits insight into the effects of Lorentz violation 
on radiation.
Note also that 
the effective permittivity \rf{perm}
depends on the frequency $p^0$ and wave vector $\vec p$,
which implies a nonlocal connection between $\vec D (x)$
and $\vec E(x)$.

The extended Maxwell equations for this analogy
directly yield 
\beq
\vec p \cdot \vec B = \vec p \cdot \vec D = 
\vec E \cdot \vec B = \vec D \cdot \vec B = 0 
\quad .
\label{perp}
\eeq
The natural right-handed triad of orthonormal vectors 
describing the vibration of the electromagnetic field
is therefore $(\hat p, \hat D, \hat B)$.
Unlike the case of conventional vacuum radiation,
the electric-field vector $\vec E$
here is orthogonal only to $\hat B$
and so lies off-axis in the $\hat p$-$\hat D$ plane.
In this analogy,
the energy density is typically transported 
neither in the direction $\hat p$
nor in the direction $\hat E \times \hat B$. 

It is useful to introduce 
a generalized refractive index $n(p)$ by
$n(p) \equiv |\vec p| / p^0$.
Its inverse is the magnitude
of the phase velocity of the mode \rf{momspaceF}.
Using the extended Maxwell equations,
we can then deduce the result 
\beq
\vec D = 
n^2 \left[ \vec E - (\hat p \cdot \vec E)\hat p \right]
\equiv n^2 \vec E_\perp
\quad ,
\label{Eperp}
\eeq
which determines the effective displacement current
directly in terms of the electric field and the momentum.
Eliminating $\vec D$ via Eq.\ \rf{Ddef}
produces a set of three linear equations 
of the form \rf{amp} for $\vec E$,
with the matrix $M$ now given by
\beq
M^{jk} \equiv n^2 \de^{jk} - \fr{p^jp^k}{(p^0)^2} - \ep^{jk}
\quad .
\label{Mrefind}
\eeq
A nontrivial solution exists if $\det(M^{jk})= 0$.
We have explicitly verified that this condition is 
equivalent to the condition of the vanishing of the determinant
of the matrix $M^{jk}$
appearing in Eq.\ \rf{Mmatrix} in subsection IV B.

In conventional crystal optics,
the permittivity is often diagonalized:
$\ep^{jk} \equiv \ep^j \de^{jk}$ (no sum),
where the eigenvalues $\ep^j$ are real.
This means that the coordinate axes are identified 
with the principle dielectric axes,
which typically represents a different coordinate system
than the special one with $\vec p = (0,0,p)$
used in section IV B.
A diagonalization of this type 
is also possible in the present analogy
because the effective permittivity is hermitian.
Substitution into Eq.\ \rf{Mrefind}
and expansion of the determinant
then produces an expression with the form 
of the Fresnel equation of crystal optics.
The sixth-order terms in the determinant cancel,
ultimately by virtue of the existence of only two independent
degrees of freedom in $\vec E$.
Solving the determinant condition
provides the dispersion relations
for the independent degrees of freedom.
If $\hat p$ is given,
the condition specifies
$p^0$ as a function of $|\vec p|$.
If $p^0$ is given,
the condition specifies $|\vec p|$
as a function of $\hat p$.

The special choice of coordinate system in subsection IV B
permits a direct demonstration with this analogy  
that for a given momentum $\vec p$ the solutions 
for the effective displacement current $\vec D$
are linearly polarized.
First,
substitute for $E^j = (\ep^{-1})^{jk} D^k$ 
in Eq.\ \rf{Eperp}.
Using observer rotation invariance 
to select a frame in which $\vec p \equiv (0,0,p)$
then yields the two simultaneous equations 
\beq
\left( \de^{ab} - n^2 (\ep^{-1})^{ab} \right) D^b = 0
\quad ,
\label{2cptD}
\eeq
where $a,b = 1,2$.
The vanishing of the determinant of the expression in parentheses
generates the analogue of the Fresnel equation
in this special coordinate system.
It can be seen directly from Eq.\ \rf{2cptD}
that for fixed $\vec p$ 
the $\vec D$ vectors for each of the two values of $n$
must lie along the principal axes of symmetry
of the two-dimensional matrix $(\ep^{-1})^{ab}$.
These two $\vec D$ vectors are perpendicular,
so an electromagnetic wave corresponding to either one is
necessarily linearly polarized.

Many other concepts of crystal optics
can be applied in the context of this analogy,
including the wave-vector and ray surfaces
and the Fresnel and other ellipsoids.
The presence of Lorentz violation 
means that the vacuum as experienced by an electromagnetic wave 
behaves like a special kind of crystal.
Our results show that 
the effective medium is optically anisotropic and gyrotropic
and exhibits spatial dispersion of the axes.
The earliest mention of effects of this type
appeared in an 1878 paper of Lorentz
\cite{1878},
and they are now well established in a variety of physical systems 
\cite{bwll}.
Thus,
the momentum dependence of the effective permittivity
corresponds to spatial dispersion of the axes.
A nonzero $k_{AF}$ produces a contribution 
to the effective permittivity 
analogous to the effects of natural optical activity
in a gyrotropic crystal,
while a nonzero $k_F$ produces effects analogous to 
spatial dispersion in 
an optically inactive and anisotropic crystal.
Partly on the basis of the hermiticity 
of the effective permittivity,
we also anticipate that in the presence of Lorentz violation
the vacuum behaves like a transparent (nonabsorptive) medium,
although a complete and elegant demonstration of this 
remains an open issue. 

The above analyses partially simplify 
if certain components of $k_F$ and $k_{AF}$ vanish.
Suppose for definiteness that $k_{AF}$ indeed vanishes 
and that the only nonzero components of $k_F$ are
$(k_F)^{0j0k}$
and components related to these by the symmetries of $k_F$.
This makes the effective permittivity real and
independent of $p^\mu$:
$\ep^{jk} = \de^{jk} - 2 (k_F)^{0j0k}$.
It is then possible,
for example,
to solve explicitly for the behavior 
if $\vec E$ is specified,
which provides yet another approach to the physics.
In this case,
$\vec E_\perp$ is the component of $\vec E$ perpendicular 
to $\hat p$ in the $\hat E$-$\hat p$ plane.
It follows that 
$\vec E_\perp = (\vec E \cdot \hat D) \hat D$,
from which one can derive
\beq
\hat p = \fr
{ (\vec D)^2 \vec E - (\vec E \cdot \vec D) \vec D }
{ \sqrt{ 
\left [ (\vec E)^2 (\vec D)^2 
- (\vec E \cdot \vec D)^2 \right ]
(\vec D)^2 }}
\quad 
\label{hatp}
\eeq
provided $\vec D$ and $\vec E$ are not parallel.
The phase velocity is given by
\beq
v_p = 1/n = \sqrt{(\vec E \cdot \vec D) / (\vec D)^2 }
\quad .
\label{phase}
\eeq
For instance,
if $\vec E = (E,0,0)$ then to lowest order in $k_F$ we find 
$\vec D \approx |\vec E| 
(1 - 2 (k_F)^{0101}, - 2 (k_F)^{0201}, - 2 (k_F)^{0301})$
and $v_p \approx 1 + (k_F)^{0101}$,
which 
in the appropriate limit 
agrees with the result for the extraordinary mode
of the example at the end of subsection IV B.
Even in this relatively simple case,
Eq.\ \rf{hatp} shows that the vector $\vec p$ 
can have a complicated structure
with components in all three directions.

The above analysis 
uses the notion of the phase velocity $v_p$.
However,
even in conventional electrodynamics
there are numerous possible definitions 
of the velocity $v$ of an electromagnetic wave,
including among others the group velocity,
the velocity of energy-momentum transport,
and the signal velocity 
\cite{sb}.
The Lorentz violation adds further complications
to this situation.
In the remainder of this subsection,
we comment on some aspects of this issue.

An important feature is that the fundamental physical constant
$c=1$ relating the space and time components of the metric
is unaffected by the Lorentz violation.
The underlying spacetime structure of the theory
is the usual one because
the apparent Lorentz breaking 
at the level of the standard model
is merely a reflection of the presence 
of nonzero tensor expectation values
in a fundamental theory with Lorentz-covariant dynamics.
Indeed, $c$ is an invariant under both observer and particle boosts.
However, the physical velocity of an electromagnetic wave 
\it can \rm be affected.
The situation is analogous to that of a fermion mass parameter $m$
in the lagrangian for the standard-model extension: 
although $m$ remains unchanged,
the physical rest mass of a particle can be affected
\cite{cksm}.

As in conventional electrodynamics,
the various definitions of physical velocity 
are inequivalent in general.
Any choice for the physical velocity $v$ 
typically differs from $c$,
although $v\approx c$ 
since $k_F$ and $k_{AF}$ are small.
The analyses above indicate that,
for any given definition,
the magnitude and direction of the velocity 
of an electromagnetic wave
can vary with the wave-vector orientation 
and the polarization.
Incidentally,
conventional crystal-optics experiments suggest that there is
no general condition requiring the velocity 
of one type of polarization to exceed the other.
For example,
the indices of refraction 
$n_o$ and $n_e$ for the ordinary and extraordinary rays,
respectively,
of the sodium D line are 
measured to be $n_o= 1.658 > n_e=1.486$ in calcspar
but are $n_o= 1.544 < n_e=1.553$ in quartz
\cite{CRC}.

For the special case involving Eq.\ \rf{phase},
$v_p$ may exceed $c$ if the sign of $k_F$ is appropriate.
In conventional electrodynamics,
a phase velocity exceeding $c$ is known to occur
in numerous physical situations,
for example,
for TE and TM modes in wave guides.
Indeed,
both the phase and group velocities can simultaneously 
exceed $c$ in certain refractive materials.
For the present theory,
it is an open issue to demonstrate that 
a phase velocity exceeding $c$
is compatible with microcausality.
It is possible in principle
that only certain sign choices for the components of $k_F$ 
lead to physically acceptable microcausal theories.
If this occurs,
it would be analogous to the usual requirement of
a particular sign for the mass-squared term 
in a (stable) scalar field theory.
In any event,
a satisfactory proof of microcausality would involve
a complete treatment at the level of quantum field theory
and lies beyond the scope of the present work. 

Another issue involving the physical velocity $v$
of an electromagnetic wave is 
its behavior under Lorentz transformations.
Since $c$ is invariant under an observer Lorentz transformation
whereas $k_{AF}$ and $k_F$ change,
$v$ is expected to transform
along with the frequency and wavelength.
This is unlike the conventional case
and is a consequence of the presence of the
background expectation values.
In contrast,
a particle Lorentz transformation,
which for a fixed polarization mode
involves remaining in the specified observer frame
but changing $\vec p$,
has no effect on $c$, $k_{AF}$, or $k_F$.
Note that if $\vec p$ is changed 
while the polarization is fixed,
the above analyses show that the frequency $p^0$ also
changes in this case.
One might instead countenance another kind of boost
in which $\vec p$ is changed but $p^0$ is unaffected,
in which case the polarization must also change.

\subsection{Constraints from Birefringence}

The existence of distinct dispersion relations for 
the independent polarizations means that birefringence is
a major feature of the behavior of an electromagnetic wave
in vacuum in the presence of Lorentz violation.
In this subsection,
we investigate some of the theoretical and experimental
implications of a birefringent vacuum.

For definiteness,
consider a monochromatic electromagnetic wave 
of frequency $p^0$. 
The electric field $\vec E (t, \vec x)$ of this wave
is formed in general from 
two independent polarization components:
\bea
\vec E (t, \vec x) &=& 
\left[\vec E_1 (\vec p_1) 
\exp(i \vec p_1 \cdot \vec x) 
\right.
\nonumber \\
&& 
\left.
\quad 
+\vec E_2 (\vec p_2) 
\exp(i \vec p_2 \cdot \vec x)\right]
\exp(- i p^0 t)
\quad .
\label{monoE}
\eea
The wave vectors $\vec p_1$ and $\vec p_2$
must satisfy the appropriate dispersion relations
for the specified frequency $p^0$.
Note that the direction of wave propagation 
must also be specified to fix completely the solution.
One possible determining method could be to require
that both component waves propagate their energy density
in a given direction.

Since the Lorentz violation is small,
we expect
$\vec p_2 = \vec p_1 + \de \vec p$,
where $\de \vec p$ is small relative to 
$\vec p_1 \approx \vec p_2$.
Substitution gives 
\bea
\vec E (t, \vec x) & \approx & 
\left[\vec E_1 (\vec p_1) 
+ \vec E_2 (\vec p_2) 
\exp(i \de\vec p \cdot \vec x)\right]
\nonumber \\ && 
\quad \qquad\qquad 
\times
\exp(- i p^0 t + i \vec p_1 \cdot \vec x) 
\quad .
\label{monoEapp}
\eea
This equation shows that the 
birefringence length scale is $|\de \vec p|^{-1}$,
which is large when $|\de\vec p|$ is small.
Since $\de\vec p$ has dimensions of mass
and since it vanishes in the absence of Lorentz violation,
its dominant terms are expected 
to be controlled by $k_{AF}$,
by a product of components of $\vec p$ and $k_F$, 
or by some combination of the two.
This is in agreement with the discussion
of the dispersion relations in previous subsections.
Note that the associated phase shift
$\De\ph \equiv \de\vec p\cdot \vec x$
cannot correctly be regarded as a phase difference between
two circular-polarization modes because typically
no such modes exist as solutions of the dispersion relations. 

In the remainder of this subsection,
we consider possible bounds on $k_F$ and $k_{AF}$
from some terrestrial, solar system, astrophysical, 
and cosmological experiments. 

First, 
we summarize the case of nonzero $k_{AF}$ but zero $k_F$.
A term of the form \rf{cptqed} appears to have been 
introduced independently on several occasions,
including among others in Ref.\ \cite{ni}
and the review \cite{np} mentioned earlier, 
although the observation that 
it is CPT violating appears to have been overlooked
prior to our earlier work \cite{cksm}.
Given the theoretical difficulties 
arising from negative contributions to the energy
as described in subsection IV B,
it seems possible that 
this term would need to be absent in nature
even if Lorentz symmetry is violated.
However,
this too is a suggestion that could be the subject of tests.

In a pioneering work
\cite{cfj},
Carroll, Field, and Jackiw 
investigated some properties of the term \rf{cptqed}
and used geomagnetic constraints
and limits on cosmological birefringence of radio waves 
to bound certain forms of the coupling coefficient $k_{AF}$.
Their treatment of geomagnetic constraints
is based on known bounds on the photon mass
\cite{photon},
and it constrains a term of the form \rf{cptqed}
with $(k_{AF})^\mu = (k,\vec 0)$ 
to $|k|\lsim 6 \times 10^{-26}$ GeV.
In contrast,
the constraints they obtain from cosmological birefringence
are considerably sharper,
primarily because the distance scales are greater.
Their investigation seeks a redshift dependence
in the established correlation
\cite{hpk}
between the intrinsic position angles
and the polarization angles
of a set of radio galaxies and quasars
at distances comparable to the Hubble length.
It constrains a particular combination of the coefficients 
for a timelike $(k_{AF})^\mu$
to $\lsim 2 \times 10^{-42}$ GeV.
A more recent analysis
\cite{nr}
claims a nonzero observed effect 
with a spacelike $(k_{AF})^\mu$
at a scale of approximately $10^{-41}$ GeV.
This has been disputed by other authors 
\cite{misc}.

We next consider the case with $k_F \neq 0$ but $k_{AF} = 0$.
In its general form,
this possibility appears to have been largely disregarded 
in the literature.
However,
the rotationally invariant term 
of the form $\half \al (\vec B^2 + \vec E^2)$ 
in the extended-QED lagrangian \rf{totaleb}
has been considered by several authors,
usually in the rescaled form involving only $\vec B^2$. 
In particular,
this term has a counterpart in the TH$\ep\mu$ formalism
\cite{llee}.
This formalism is a phenomenological parametrization
for the motion and electromagnetic interactions
of charged pointlike test particles
in an external spherically symmetric and static gravitational field.
It has been extensively used for quantitative tests 
of the foundations of gravity,
including local Lorentz invariance.
In this context,
clock-comparison experiments have constrained 
the analogue of the parameter $\al$
to better than about one part in $10^{21}$
\cite{cw}.
An improvement over this bound
of about an order of magnitude 
may be possible based on the existence 
and properties of high-energy cosmic rays
\cite{cg}.

In the general case with nonzero $k_F$
and violation of rotational invariance,
the sharpest bounds are likely to emerge once again
from observational constraints on cosmological birefringence.
However,
the discussion following Eq.\ \rf{monoEapp} shows 
there is a significant difference in the $k_F$ case:
the phase shift $\De\ph$ here depends on 
a product of components of $k_F$ and $\vec p$,
whereas in the $k_{AF}$ case 
$\De\ph$ depends only on $k_{AF}$.
This behavior can be seen explicitly,
for example,
in the special analytical solution presented
at the end of subsection IV B.

The linear dependence of $\De \ph$
on momentum or wave number 
implies an inverse dependence on wavelength.
The rotation measures and intrinsic position angles
of radio sources
\cite{hpk}
are obtained by a fitting procedure that assumes
a quadratic dependence on wavelength 
(proportional to the rotation measure 
and attributed to Faraday rotation)
with a wavelength-independent zero offset
(the intrinsic position angle).
This procedure is suitable for obtaining constraints 
on $k_{AF}$,
which would generate an extra wavelength-independent effect,
but may be inadequate to place a reliable bound on $k_F$
or to detect the associated wavelength-dependent effects.
It therefore appears somewhat involved 
to obtain an accurate estimate 
of the constraints on $k_F$
from cosmological birefringence.

Although a complete treatment lies outside our present scope,
a crude estimate of an attainable bound on $k_F$ can readily
be found.
It is plausible to 
suppose that the results of a careful analysis
would provide a limit on a product of 
certain components of $k_F$ and $\vec p$ 
comparable to that of order $10^{-42}$ GeV
obtained for $k_{AF}$ in Ref.\ \cite{cfj}.
The radio sources typically involve wavelengths
of order 10 cm,
which corresponds to an inverse wavelength 
of about $10^{-15}$ GeV.
This suggests that an upper bound 
of approximately $10^{-27}$ could be placed 
on at least some of the 
(dimensionless) coefficients $k_F$. 
The tightness of this constraint 
and the apparent feasibility of the analysis 
suggests this investigation would be worthwhile to pursue.
Ideally,
a complete study would obtain 
combined bounds on both of the coupling coefficients 
$k_F$ and $k_{AF}$.

An interesting implication of 
the (inverse) wavelength dependence
of the birefringence 
is that shorter wavelengths are more sensitive 
to the effects.
Although it may be infeasible in practice,
a measurement of cosmological birefringence 
comparable to the above but obtained with,
say,
optical sources would be much more sensitive 
to possible effects from $k_F$.
Optical wavelengths are a factor of about $10^{-6}$ 
of radio wavelengths,
which would correspond to a millionfold improvement 
in sensitivity to $k_F$.

Other bounds on Lorentz violation could be deduced. 
In the next section,
we show that one-loop radiative corrections
induce a dependence of $k_F$ on the coefficients $c_{\mu\nu}$
in Eq.\ \rf{lorviolqed}
for the extended QED.
This suggests that if a tight bound were obtained on $k_F$
as above,
an indirect constraint might also be inferred 
on $c_{\mu\nu}$.
The latter constraint would be weaker 
by a factor of the fine-structure constant,
but the limits deduced would 
nonetheless probably be comparable to the best ones
attainable in other tests of Lorentz symmetry. 

If a nonzero effect is detected in the future,
it might be of some theoretical interest to investigate 
the possibility of a correlation 
between the particular coupling coefficients involved
and the motion of the Earth 
relative to the cosmic microwave background radiation.
The point is that the apparent Lorentz violation
induces boost (and orientation) dependence
in experiments
\cite{ak}.
Although
the standard-model extension strictly has no preferred frame,
the coupling coefficients must take a canonical form 
in some observer frame
\cite{cksm}.
If the latter is at rest
with respect to the cosmic microwave background radiation,
a small deviation from the canonical form
might arise from the Earth's motion.
Although the Earth's speed in this frame is about $10^{-3}c$,
the sensitivity of the birefringence measurements 
might nonetheless be sufficient to detect its effects.

\section{RADIATIVE CORRECTIONS}

We next examine some radiative corrections
to the pure-photon sector.
In subsection V A,
CPT-odd terms are investigated.
Of particular interest
is whether the tree-level vanishing of
the coefficient $k_{AF}$ 
in Eq.\ \rf{cptqed} for the QED extension,
which would eliminate negative contributions to the energy,
is reasonable in the light of quantum effects.
The point is that
the latter might in principle induce a nonzero coefficient 
through radiative corrections from another sector 
of the theory.
Other quantum corrections that might generate an instability
through the linear term $-(k_A)_\mu A^\mu$
are considered at the end of this subsection.
In subsection V B,
we study quantum corrections in the CPT-even sector,
involving the coefficient $k_F$.

The analysis in this section is based on the quantization
discussed in Ref.\ \cite{cksm}.
It is largely at the one-loop level
and for leading-order Lorentz-violating effects,
and it is primarily limited to issues 
involving radiative corrections to the pure-photon sector.
A few results are also presented for higher loops and effects
in other parts of the standard-model extension.

An interesting issue indirectly related to
some calculations in subsection V A 
is whether the anomaly cancellations
occurring in the conventional standard model
still hold for its extension presented in section II.
Three known types of chiral gauge anomaly are relevant 
\cite{gm}.
It lies beyond our present scope to
provide a complete analysis of all these 
in the presence of the Lorentz-violating terms,
and it is certainly conceivable that the latter 
would modify the standard derivations. 
We do expect,
however,
that the usual cancellations of the
(abelian, singlet, and nonabelian) 
triangular gauge anomalies
and the nonperturbative global SU(2) anomaly
indeed remain valid.
The point is that the standard-model extension
has the same multiplets and the same gauge structure
as the conventional case,
so the group-theoretic underpinnings of the usual analyses 
are unaffected.
The situation for the third type of anomaly,
which is the mixed gauge-gravitational chiral anomaly
associated in part with local Lorentz transformations,
is less clear.
The presence of Lorentz violations might appear to suggest
a potentially nonzero contribution to this anomaly.
However,
a careful analysis is needed because
observer Lorentz invariance is in fact maintained
in the standard-model extension.

\subsection{CPT-Odd Terms}

As discussed above,
the possible difficulties 
with negative contributions to the energy
and the tight experimental constraints 
suggest that the coefficient $k_{AF}$ vanishes.
If it is set to zero at tree level,
the issue arises as to whether it acquires
radiative corrections from quantum loop corrections.
If so, 
there could be both theoretical and experimental arguments 
suggesting associated constraints on
certain other coefficients in the standard-model extension.
The issue of the vanishing of radiative corrections
to $k_{AF}$ therefore has the potential 
to provide a nontrivial consistency check
on the theory.
In the present subsection we investigate this,
assuming that $k_{AF}$ is zero at tree level
and beginning with one-loop effects
at leading order in the Lorentz-breaking terms.
Remarkably,
as we show next,
the structure of the standard-model extension 
is such as to preserve a vanishing coefficient $k_{AF}$
at this level.

A radiative contribution to $k_{AF}$ would represent
a correction to the photon propagator.
In the standard-model extension,
the Feynman rules for leading-order effects
from Lorentz-violating terms
take the form of insertions on propagators or at vertices
already existing in the conventional theory
\cite{cksm}.
Also,
the photon interacts with charged particles as usual,
so the only possible diagrams modifying the photon propagator
at the one-loop level
are those of the standard one-loop vacuum polarization
but with an insertion either on an internal charged-particle line
or at one of the vertices.

The apparently daunting task of examining every possible 
insertion implied by the extra terms in the standard-model
extension can be simplified by taking advantage
of the discrete operations C, P, and T.
A radiative term purporting to contribute 
to the coefficient $k_{AF}$
must have appropriate transformation properties under
these discrete symmetries.
In particular,
it must be C even and PT (and CPT) odd, 
although either of the two possible combinations of P and T
could occur.
At the level of the QED extension with electrons and positrons,
the only term of this type
is the one with coupling coefficient
$b_\mu$ in Eq.\ \rf{cptvioleqed}.
This is true even if the discarded linear term 
$-(k_A)_\ka A^\ka$ in the lagrangian were present, 
which is C and CPT odd.
At any loop order,
contributions must therefore involve
an odd number of line insertions arising 
from the term with coefficient $b_\mu$.
At the one-loop level in the full standard-model extension,
similar terms involving the other lepton and quark fields
would also contribute to appropriate internal lines.
However,
only one additional distinct type 
of one-loop contribution appears, 
involving a vertex correction proportional to 
the coefficient $(k_2)_\mu$ in Eq.\ \rf{cptgauge}
in a diagram with a $W^+$-$W^-$ loop. 
The demonstration that no net contributions to $k_{AF}$ 
arise at one loop therefore involves
consideration of only terms involving
the $b_\mu$-type and the $(k_2)_\mu$ coefficients.

Excluding the external photon legs,
any contributions to the vacuum polarization must have 
dimensions of mass squared.
The leading-order contribution
to $k_{AF}$ must involve 
both a momentum factor from the necessary derivative 
on an external leg 
and one power of either $b_\mu$ or $(k_2)_\mu$.
Since these factors already give the correct dimensionality,
any others must appear in dimensionless
combinations of the photon momentum $p^\mu$
and the mass $m$ of the particle in the loop.
This is confirmed by the explicit calculation below.

We first consider corrections to the one-loop vacuum polarization 
involving $b_\mu$.
Each such two-point diagram has the usual form
except for an insertion of the factor $-ib_\mu \ga_5\ga^\mu$
on one internal fermion line.
{}From the perspective of the fundamental theory,
a one-loop two-point diagram with a fermion-line insertion 
is closely related to a one-loop three-point diagram
containing the same two photon legs
together with a third leg involving a coupling
to an axial vector.
A fermion-line insertion in the two-point diagram
can then be viewed as a limit of this three-point diagram
in which there is zero momentum transfer 
to the axial-vector leg and the axial vector 
is replaced with a vacuum expectation value.

This line of reasoning is interesting 
because a one-loop three-point diagram
with an axial-vector and two photon couplings
is directly related to a triangular gauge anomaly.
If the axial vector is a gauge field in the underlying theory,
such anomalies must cancel
for the theory to be renormalizable.
One might therefore conjecture that 
the cancellation of these anomalies
could also imply cancellation of the limiting 
two-point diagrams in the standard-model extension.
If true,
this provides another link relating consistency 
of the standard-model extension
to the spontaneous nature of the Lorentz violation
in the underlying theory.
Next,
we develop a line of reasoning that
provides insight into this question.

Independently of the issue of corrections to 
the photon propagator,
the requirement that the triangular anomalies cancel
in the underlying theory implies a constraint 
on coefficients of the type $b_\mu$
that is of interest in its own right.
It turns out that this constraint is relevant 
to the photon propagator,
so we begin by deriving it.

Consider first the origin in the underlying theory
of the axial-vector coupling in the triangle diagram.
Prior to spontaneous symmetry breaking,
the fundamental lagrangian may contain several terms 
of the general form
$g^a_\ps (T^a)_{\mu\nu\ldots\rh}\overline{\ps} 
(\Ga^a)^{\mu\nu\ldots\rh}{\ps}$
for each fermion species $\ps$,
where $T^a$ is a tensor field,
$\Ga^a$ is a gamma-matrix structure,
$g^a_\ps$ is the associated coupling constant,
and $a$ is a label ranging over the set of tensor fields 
that couple to the species $\ps$.
Note that the only acceptable line insertions
in the two-point one-loop diagram are flavor diagonal,
so in the present context
contributions are possible only from the diagonal components
of Eqs.\ \rf{cptvioll} and \rf{cptviolq}.
We therefore disregard 
possible cross-couplings between fermion species
in this derivation.

Each of these lagrangian terms
can be decomposed in terms of the usual
16 basis gamma matrices in four dimensions.
Collecting terms produces for each fermion species 
a lagrangian separated into five parts,
one for each of the five types of fermion bilinear:
scalar, pseudoscalar, vector, axial-vector, and tensor.
The particular components of the fields $T^a$ 
that multiply the axial-vector bilinear
can be regarded as a set of effective axial vectors
$A^a_{5\mu}$
with associated coupling constants $g^a_\ps$.
These axial vectors are the fields
relevant for the one-loop three-point diagrams of interest.
When the axial vectors  
$A^a_{5\mu}$ acquire vacuum expectation values
$\vev{A^a_{5\mu}}$,
their net contribution generates
the coupling coefficient 
$b^\ps_\mu \equiv \sum_a g^a_\ps \vev{A^a_{5\mu}}$
at the level of the standard-model extension
for this species of fermion.

The triangle diagram for one axial vector $A^a_{5\mu}$
and two photons $A_\mu$ 
has an anomaly proportional to the product of 
$g^a_\ps$ and $q_\ps^2$,
where the latter is the charge of the fermion $\ps$.
When summed over all fermion species,
the anomaly-cancellation condition is therefore
\beq
\sum_\ps q_\ps^2 g^a_\ps = 0
\quad
\label{anomcanc}
\eeq
for each $a$.
Multiplying this equation by
$\vev{A^a_{5\mu}}$
and summing over $a$ yields the constraint
\beq
\sum_\ps q_\ps^2 b^\ps_\mu = 0
\quad 
\label{bconstraint}
\eeq
on coupling coefficients of the $b_\mu$ type. 
Note that,
at the level of the standard-model extension,
the sum over all fermion species 
would include the leptons and the quarks.
Also,
in contrast to the usual anomaly-cancellation mechanism
which produces a single condition,
Eq.\ \rf{bconstraint} is a set of four constraints.
This is a direct consequence of spontaneous Lorentz breaking,
in which for each $a$ the vacuum expectation value 
$\vev{A^a_{5\mu}}$ involves four numbers. 

Next,
we present the results of an explicit calculation 
of the $b_\mu$-linear one-loop corrections 
to the photon propagator 
involving a fermion of mass $m$ and charge $q$.
There are two diagrams to consider,
since a factor $-ib_\mu \ga_5\ga^\mu$
can be inserted on either of the two internal lines.
Using an argument similar to the standard one 
proving the Furry theorem,
the two diagrams can be shown to give identical contributions
to the amplitude.
Omitting the external photon legs,
the correction to the two-point amplitude 
for a photon of four-momentum $p^\mu$ then becomes 
\bea
&&\overline{\om}^{\mu \nu}(p,m,b) = 
- 2 i q^2 b_\la
\nonumber\\ &&
\quad
\times \int \fr {d^4 l} {(2 \pi)^4}
{\rm Tr} 
\left[
\ga^{\mu} S_F (l - p) \ga^{\nu} S_F(l) \ga_5 \ga^\la S_F(l)
\right] ,
\label{omamp}
\eea
where $l^\mu$ is the momentum of the fermion in the loop
and $S_F(l) = i (\not l - m + i\ep)^{-1}$ 
is the usual fermion propagator. 

As anticipated above,
the expression \rf{omamp} is related to one appearing
in the calculation of the triangular gauge anomaly.
It can directly be verified that
\beq
\overline{\om}^{\mu \nu}(p,m,b) 
\equiv q^2 b_\la T^{\mu\nu\la}(-p,p)
\quad ,
\label{omamprel}
\eeq
where $T^{\mu\nu\la}(p_1, p_2)$
is the standard amplitude for the triangle diagram 
with one axial-vector coupling in conventional QED.
The full anomaly amplitude $T^{\mu\nu\la}(p_1, p_2)$
can be regularized in the Pauli-Villars scheme
and reduced to a set of integral expressions 
\cite{lr,sa}.
These can be evaluated in closed form 
for the present case of interest.
For $p^2<4m^2$,
we find
\widetext
\top{-2.8cm}
\hglue -1 cm
\beq
\overline{\om}^{\mu \nu}(p,m,b) 
= \fr{q^2 b_\la}{2\pi^2}
p_\ka \ep^{\ka\la\mu\nu}
\Biggl[
1 - \fr 4 {\sqrt{(p^2/m^2)(4 - p^2/m^2)}}
\tan^{-1} 
\left( \sqrt{\fr{p^2/m^2}{4 - p^2/m^2}} \right) 
\Biggr]
\quad .
\label{omampexp}
\eeq
\bottom{-2.7cm} 
\narrowtext 
\noindent
Note that this expression is gauge invariant, 
\beq
p_{\mu} \overline{\om}^{\mu \nu} = 
p_{\nu} \overline{\om}^{\mu \nu} = 0
\quad ,
\label{gauginv}
\eeq
as expected.

At this stage,
the issue of radiative corrections to $k_{AF}$
can be addressed.
The result \rf{omampexp} is finite.
Since no divergence cancellation is necessary,
a zero value of $k_{AF}$ at tree level is
consistent with a renormalizable theory.
Moreover,
$\overline{\om}^{\mu \nu}$ vanishes 
for the on-shell condition $p^2 = 0$,
as is to be expected in a renormalizable theory
without a radiatively induced phase transition.
Thus,
none of the finite radiative corrections 
have the form needed to modify the coefficient $k_{AF}$,
and they are therefore irrelevant to the analysis 
of cosmic birefringence in subsection IV D.

The above results might make it seem
tempting to conclude that 
there are no $b_\mu$-linear one-loop radiative corrections  
affecting $k_F$.
However,
such a conclusion would be premature.
The integral $T^{\mu\nu\la}(-p,p)$
in Eq.\ \rf{omamprel} 
is superficially linearly divergent.
As usual,
this introduces an ambiguity 
because a shift in the loop momentum $l^\mu$
produces a shift in the value of the integral:
$T^{\mu\nu\la}(-p,p) \to 
T^{\mu\nu\la}(-p,p) 
+ \ze p_\ka \ep^{\ka\la\mu\nu}$,
where $\ze$ is a constant.
Certain choices of regularization scheme
could therefore generate an additional term of the form 
\beq
\de \overline{\om}^{\mu \nu}(p,b) 
= \ze q^2 b_\la p_\ka \ep^{\ka\la\mu\nu}
\quad
\label{zedef}
\eeq
to the result \rf{omampexp},
which would represent a 
regularization-dependent
radiative correction to $k_{AF}$.
Note that this does not occur in the Pauli-Villars scheme
because the term \rf{zedef} is mass independent,
so in this case the regularization
automatically subtracts it.

Ambiguities of the general form \rf{zedef}
involving combinations of the external momenta
arise in the standard triangular-anomaly diagram
with a finite momentum transfer from the two photons
to the axial vector.
In this case,
the ambiguity is conventionally fixed 
by imposing U(1) current conservation.
However,
in the general context the presence of an ambiguity 
is independent of the issue of anomalies.
Under certain circumstances,
anomalies can appear in superficially convergent
(nonabelian pentagon) diagrams that are ambiguity-free
\cite{wb}.
Also,
ambiguities originating in loop-momentum shifts 
for divergent amplitudes
other than those associated with anomalies
are a standard feature of quantum field theories.
For example,
the usual vacuum-polarization diagram has an ambiguity.
Similarly,
results such as the Furry theorem
rely on a consistent assignment of loop momenta.
Typically,
these ambiguities either appear as finite constant modifications
to divergent constants
or can be eliminated by imposing gauge invariance.

A striking feature of the $\ze$ ambiguity
is that it arises without an associated divergence
and is gauge invariant.
It therefore cannot be fixed by the usual methods.
Thus,
gauge invariance ensures the Ward identities are satisfied,
so vector-current conservation holds for any $\ze$.
Also,
the ambiguity fails to produce an anomaly 
in the axial-current conservation law
because there is zero momentum transfer
away from the loop at the axial vertex for any $\ze$.
However,
the mass independence of the term \rf{zedef}
implies that the fermion mass circulating in the loop 
could in principle be arbitrarily large
without affecting the value of $\ze$,
which intuitively seems unphysical
and would appear to suggest that $\ze$ must vanish. 

In the standard triangular-anomaly diagram,
fixing the ambiguity 
by requiring vector-current conservation
places the anomaly in the axial Ward identity.
If the axial vector is ungauged,
chiral-current conservation is then violated
and the anomaly may have physical consequences.
An example of this occurs in the decay $\pi \to 2\ga$.
If instead the axial vector is a gauge field,
then the anomaly destroys renormalizability 
unless the total anomaly contribution 
from all fermion species vanishes.
A cancellation of this type,
which is widely used in model building,
implicitly assumes the
ambiguity has been fixed in a standard way
in all contributing diagrams.
This could be regarded as a (reasonable) choice 
made to obtain a satisfactory theory.

If a similar choice is made for the present case,
so that the same regularization scheme is adopted for
all the contributing diagrams  
and therefore the same constant $\ze$ appears in each,
then it can be argued that the anomaly cancellation
in the underlying theory
causes the ambiguity to disappear.
Thus,
suppose as above we assume gauged axial vectors
in a renormalizable underlying theory,
so that the anomaly-cancellation condition
\rf{bconstraint} must hold.
Then,
the net contribution to the photon propagator 
from the ambiguous terms is given by 
$\sum_\ps \ze q_\ps^2 b^\ps_\la 
p_\ka \ep^{\ka\la\mu\nu}$,
which vanishes by Eq.\ \rf{bconstraint}.
This confirms the conjecture made in the first
part of this subsection:
the anomaly cancellation implies the absence of 
$b_\mu$-linear one-loop radiative corrections
to $k_{AF}$.

Note that this argument presupposes that 
the axial vectors $A^a_{5\mu}$ are gauged and 
that a consistent choice of regularization is used.
To demonstrate the absence of negative-energy 
contributions to the theory at this level, 
it suffices that a natural procedure of this kind exists.
If an ambiguity $\ze$ did remain in the theory,
it would seem to suggest that at the quantum level 
there would be a spectrum of physically allowed theories.
The issue of determining the correct one
would then become experimental,
much as the values of the renormalized couplings and masses
are experimentally determined.
However,
in the present case there 
are both theoretical and experimental 
reasons to believe that $\ze$ vanishes.

We next address some issues 
arising in higher perturbation orders.
Consider first the case of 
the photon propagator in the extended QED.
At any loop order 
but with only one CPT-violating insertion of $b_\mu$,
all diagrams are superficially divergent
and hence can be expected to have ambiguities.
In parallel with the previous case,
these diagrams can be related to
higher-loop three-point triangle diagrams
with one axial-vector and two photons
on the external legs.
The Adler-Bardeen theorem 
\cite{ab}
shows that the anomalies arising
from the one-loop triangle diagram
are unaffected at higher loops.
This implies that the constraint \rf{bconstraint}
holds at arbitrary loop order.
However,
it follows as before that the total ambiguity
is proportional to this constraint
and so vanishes.
If this argument holds,
then there can be no $b_\mu$-linear contributions to $k_{AF}$
at any order in the fine-structure constant.

Diagrams that involve higher-order Lorentz violation
may also be of potential theoretical importance.
Their transformation properties under discrete symmetries
place strong constraints on their possible contributions
to $k_{AF}$,
as in the lowest-order case.
For example,
in the extended QED at the quadratic level of Lorentz violation,
only a product of the coefficients $b_\mu$ and $c_{\mu\nu}$
can appear.
At the one-loop level,
all higher-order diagrams are related 
to polygonal diagrams in the underlying theory
that couple two photons to a variety of vector, axial-vector,
and tensor fields.
At least one factor of $b_\mu$ is required,
so a chiral coupling must be involved
and a cancellation mechanism may still apply.
The implication of the consistency of the underlying theory
for corrections to $k_{AF}$
at all orders in Lorentz violation 
and including possible higher-loop corrections
remains an open issue.
We remark,
however,
that the effects at the cubic levels and above
are at most of theoretical interest
as they would be well below experimental detection
for the levels of Lorentz violation considered 
in the present work.

At the level of the standard-model extension,
a possible lowest-order one-loop correction
to the photon propagator
could in principle also arise 
from the coefficient $k_2$
when a $W^+$-$W^-$ pair circulates in the loop.
Indeed,
there would be a contribution from insertions on the 
gauge-boson lines
and another,
related to the first by gauge invariance,
involving a modified vertex.
However,
if the term involving $k_2$ were to exist,
it would exhibit difficulties with negative
contributions to the energy,
as does $k_{AF}$.
One option is therefore that $k_2$ vanishes at tree level,
which eliminates the possible associated radiative corrections
to the term involving $k_{AF}$. 
An issue then arises concerning possible radiatively induced 
contributions to $k_2$ from the fermion sector.
We conjecture that,
if all the terms $k_0$, $k_1$, $k_2$, $k_3$ 
vanish at tree level,
then no radiative corrections to any of these coefficients
arise from the fermion sector. 
An anomaly-cancellation mechanism
would again play a role,
although nonabelian fields would now be involved
and so the singlet and nonabelian anomalies 
would also be relevant.
The situation in the standard-model extension 
at higher loops remains open.

Finally,
we present a few remarks 
about the possibility of radiative corrections
to a hypothetical linear term
of the form $- (k_A)_\ka A^\ka$
in the absence of a photon mass.
This term is C and CPT odd.
In the extended QED,
only terms of the type $a_\mu$
have this symmetry.
As discussed in section III,
a field redefinition can be used to eliminate
these (flavor-diagonal) terms,
and hence they are unobservable in any experiment.
Since the electromagnetic interactions are C even,
at lowest order in Lorentz-violating coefficients
there cannot be any radiative corrections to $k_A$ 
at \it any \rm order in QED loops.
Any contributions that might arise
at higher orders in Lorentz violation
would again be related to polygonal diagrams 
in the underlying theory.
It would be of some theoretical interest to 
investigate the possible contributions to these terms.

\subsection{CPT-Even Terms}

In contrast to the situation for the CPT-odd terms,
a nonzero tree-level value for the CPT-even term \rf{lorqed} 
with coefficient $k_F$
presents no immediate theoretical difficulty.
We have shown in section IV D
that it is experimentally feasible to place
relatively tight bounds on $k_F$
from measurements of cosmological birefringence,
although this has not yet been done
and the wavelength dependence may result
in constraints somewhat weaker than those on $k_{AF}$.
Nonetheless,
the attainable limits on $k_F$ are of interest
because they might in principle be sufficiently sharp 
to be sensitive to effects at a scale comparable
to finite radiative corrections from the fermion sector.
It is therefore of interest to determine whether
the coupling $k_F$ must be present for renormalizability
and, if so,
which fermion-sector coupling coefficients are involved.
In this subsection,
we investigate this issue
in the context of the extended QED.

At the one-loop level and to leading order in Lorentz violation,
the possible radiative contributions to 
the coefficient $k_F$ in the term \rf{lorqed} 
are significantly constrained
by the requirements of discrete symmetries.
This term 
is both C and CPT even,
and an inspection shows that the only other type of term  
in the fermion sector with these properties
is the term with coefficient $c_{\mu\nu}$ 
in Eq.\ \rf{lorvioleqed}.
It contributes both on the loop
through fermion-line insertions with a derivative
and at the vertices 
through the extra gauge coupling.

The form of the $c_{\mu\nu}$-linear correction 
$\overline{\om}^{\mu\nu}(p,m,c)$
to the two-point amplitude
for a photon of four-momentum $p^\mu$
is strongly constrained by its discrete-transformation properties,
observer Lorentz covariance,
and the requirements \rf{gauginv} of gauge invariance.
Thus,
invariance under CPT implies 
$\overline{\om}^{\mu\nu}(p,m,c)$
is an even function of $p^\mu$.
Also,
by virtue of the definition of the photon propagator
as a vacuum expectation value of a time-ordered product,
$\overline{\om}^{\mu\nu}(p,m,c)$
is symmetric under the combination of a sign change
$p^\la \rightarrow -p^\la$ of the momentum 
and an interchange $\mu \leftrightarrow \nu$ 
of the spacetime indices.
These conditions imply that
the correction to the photon propagator
at any order in the fine-structure constant
but at linear order in $c_{\mu\nu}$
must take the form
\bea
\overline{\om}^{\mu \nu}(p,m,c) & = & 
i c_{\al \be}\left( 
A g^{\al \be} (p^2 g^{\mu \nu} - p^{\mu} p^{\nu}) 
\pt{\fr D {m^2} }
\right. 
\nonumber \\ 
& & 
\left. 
+ B \left[(p^2 g^{\al \mu} g^{\be \nu} 
     - g^{\al \mu} p^{\be} p^{\nu}
     - g^{\al \nu} p^{\be} p ^{\mu})
\right. 
\right. 
\nonumber \\ 
& & 
\left. 
\left. 
\qquad \qquad \qquad \qquad \qquad
  + (\al \leftrightarrow \be)\right]
\right. 
\nonumber \\ 
& & 
\left. 
+ C g^{\mn} p^{\al} p^{\be} 
+ \fr D {m^2} p^{\al} p^{\be} p^{\mu} p^{\nu}
\right) 
\quad .
\label{gi}
\eea
Here, 
$A$, $B$, $C$, and $D$
are (possibly divergent) scalar functions of 
$p^2/m^2$
obeying the relationship
\beq
C - 2B + \fr{p^2}{m^2} D = 0
\quad 
\label{ginv}
\eeq
to ensure gauge invariance.

Some information about photon propagation 
under specified circumstances
can be deduced from Eq.\ \rf{gi}
under the assumption that the scalar functions
$A$, $B$, $C$, $D$ have been regularized
as needed and divergent contributions have been 
removed by the renormalization procedure.
For example,
in the case of cosmological birefringence
of interest in subsection IV D,
the photon momentum can be taken as on shell
and the Lorentz gauge condition can be applied.
In Eq.\ \rf{gi},
this corresponds to setting to zero both
$p^2$ and the momentum factors $p^\mu$ and $p^\nu$ 
with specific indices
$\mu$ and $\nu$.
This leaves only the term 
$c_{\al\be}C(0)g^{\mu\nu} p^\al p^\be$.
This is precisely of the form needed
for radiative corrections to the coefficient $k_F$,
which can thus be seen to be governed 
in this gauge by the on-shell value of $C$.

To obtain the explicit result
and as a check on the renormalization procedure 
when Lorentz violations are involved,
we have directly performed the one-loop calculation.
This also verifies the structure of Eq.\ \rf{gi}.
The terms in Eq.\ \rf{lorvioleqed}
associated with the coupling coefficient $c_{\mu\nu}$
lead to four new $c_{\mu\nu}$-linear
one-loop vacuum-polarization diagrams.
The possibility of fermion-line insertions
arising from the derivative coupling
leads to two diagrams,
each with one insertion on one of the two internal fermion lines.
The appearance of modified vertices
from the extra gauge coupling
leads to another two,
each with one normal and one modified vertex.
These two types of contribution are related
by gauge invariance.
Indeed,
we anticipate this gauge invariance
leads to Ward-type identities valid at arbitrary loop order,
although an explicit demonstration of this remains an open issue.

The sum of the four additional diagrams generates 
a one-loop correction to the photon propagator of
\bea
&&\overline{\om}^{\mu \nu}(p,m,c) = 
i q^2 c_{\al \be} 
\nonumber\\ &&
\qquad
\times \int \fr {d^4 l} {(2 \pi)^4}
\biggl\{
l^{\be} {\rm Tr}[\ga^{\mu} S_F (l - p) 
   \ga^{\nu} S_F(l) \ga^{\al} S_F(l)]
\nonumber \\
& & \qquad\qquad
+ (l - p)^{\be} {\rm Tr}[\ga^{\mu} S_F (l - p) 
   \ga^{\al} S_F(l - p) \ga^{\nu} S_F(l)] 
\nonumber \\
& & \qquad\qquad
- i g^{\be \mu} {\rm Tr}[\ga^{\al} S_F (l - p) \ga^{\nu} S_F(l)]
\nonumber \\
& & \qquad\qquad
- i g^{\be \nu} {\rm Tr}[\ga^{\mu} S_F (l - p) \ga^{\al} S_F(l)]
\biggr\}
\quad ,
\label{vacpolc}
\eea
where the first two terms arise from line insertions
and the last two from the modified vertices.

The integral in Eq.\ \rf{vacpolc}
is superficially quadratically divergent.
It has the standard ambiguity,
arising from the possibility of shifting the integration variable,
that is (largely) fixed by imposing gauge invariance.
The denominators arising from the fermion propagators $S_F$
can be combined with the usual Feynman parametrization.
All the necessary shifts performed 
in the resulting integration variables 
must be the same,
so that the contributions from 
the surface terms remain gauge invariant.
To accomplish this,
it is convenient to separate $\overline{\om}^{\mu\nu}$
into two pieces that can be parametrized 
so as to maintain the equivalence of shifts.

\widetext
\top{-2.8cm}
\hglue -1 cm
We define
$\overline{\om}^{\mn} =
\overline{\om}^{\mn}_{(1)} 
+ \overline{\om}^{\mn}_{(2)}$,
where
\bea
\overline{\om}^{\mn}_{(1)} 
& = & 
4 q^2 c_{\al \be} \int \fr {d^4 l}{(2 \pi)^4} \fr 1 {\De} 
\biggl\{ 
g^{\mn}
\left[ 
(l - p)^\al l^\be + (\al \leftrightarrow \be)
\right]
+\left[
g^{\mu \be}
[l \cdot (l - p) - m^2] g^{\nu \al} 
+ \left( \mu \leftrightarrow \nu \right)
\right]
\nonumber \\
& & 
\pt{ 4 q^2 c_{\al \be} \int \fr {d^4 l}{(2 \pi)^4} \fr 1 {\De} }
-\left(
\left[
g^{\mu \al}[l^\be (l - p)^\nu + (\be \leftrightarrow \nu) ]
+ \left( \mu \leftrightarrow \nu \right)
\right]
+ \left( \al \leftrightarrow \be \right)
\right)
\biggr\}
\label{om1}
\eea
and
\bea
\overline{\om}^{\mn}_{(2)} 
& = & 
8 q^2 c_{\al \be} \int \fr {d^4 l}{(2 \pi)^4} \fr 1 {\De^2} 
\biggl\{
\left( l^\al l^\be [(l - p)^2 - m^2] 
+ (l - p)^\al (l - p)^\be (l^2 - m^2)\right)
\nonumber \\
& & 
\pt{ 8 q^2 c_{\al \be} \int \fr {d^4 l}{(2 \pi)^4} \fr 1 {\De^2}} 
\times
\left( m^2 g^{\mu \nu} - l \cdot (l - p) g^{\mu \nu} 
+ (l - p)^\mu l^\nu + l^\mu (l - p)^\nu \right) \biggr\}
\quad .
\label{om2}
\eea
\bottom{-2.7cm} 
\narrowtext 
\noindent
In these expressions,
$\De = (l^2 - m^2)[(l - p)^2 - m^2]$.  
The same shift is introduced in all the integrals,
thereby preserving gauge invariance,
with the substitution in Eq.\ \rf{om1} of
\beq
\fr 1 {\De} = 
\int_{0}^{1} dz \fr 1 {[k^2 + z(1 - z) p^2 - m^2]^2}
\quad 
\label{om1sub}
\eeq
and the substitution in Eq.\ \rf{om2} of
\beq
\fr 1 {\De^2} = 
\int_{0}^{1} dz \fr {6z(1 - z)} 
{[k^2 + z(1 - z) p^2 - m^2]^4}
\quad ,
\label{om2sub}
\eeq
where $k = l - pz$ is the new integration variable.

The divergences in the resulting integrals can
be treated using dimensional regularization
in $D = 4 - \ep$ dimensions.
Performing various partial integrations,
we obtain for $p^2 < 4 m^2$
a radiative correction of the form \rf{gi} with 
\bea
A & = & - B = \om(p^2 / m^2) 
\quad , \nonumber \\
C & = & - 2 \om(p^2 / m^2) - \fr {p^2}{m^2} D 
\quad , \nonumber \\
D & = & 2 \fr {\partial} {\partial (p^2 / m^2)}
\om(p^2 / m^2) 
\quad . 
\label{ABCDresult}
\eea
Here,
$\om(p^2 / m^2)$ is the standard vacuum-polarization result,
given by 
\bea
&&\om(p^2 / m^2) =
\fr {q^2} {4 \pi^2} \left( 
\fr{1}{3}
\left(\fr{2} {\ep} - \ga \right) 
\right.
\nonumber\\&&
\qquad
\left.
- 2 \int_{0}^{1} dz ~z(1-z) 
\ln [1 - z(1-z)(p^2/m^2)]
\right) ,
\label{omusual}
\eea
where $\ga$ is the Euler constant.
Note that the results \rf{ABCDresult}
satisfy the gauge-invariance condition \rf{ginv}.

The above calculation shows that the scalar function $C$
contains a momentum-independent divergence.
As described above,
the on-shell value of $C$
determines the coefficient $k_F$ in Lorentz gauge,
so the appearance of this divergence shows 
that a bare coefficient $k_F$ must be present 
in the original theory for renormalizability.
The renormalization procedure then
removes the infinite and (ambiguous) constant pieces,
leaving a physical coefficient $k_F$
(to be determined by experiment)
and a set of finite radiative corrections 
governed by the ratio $p^2 / m^2$.

We have seen in subsection IV D
that a nonzero value of the coefficient $k_F$
induces cosmological birefringence.
The above calculation shows that 
imposing a zero value of this coefficient
at tree level is incompatible with renormalizability.
It is therefore reasonable to expect a nonzero 
physical value of $k_F$.
Although nonrigorous,
a heuristic argument might also be used to
provide a relationship between the physical values of 
$k_F$ and $c_{\mu\nu}$:
for consistency of perturbation theory,
it is plausible that 
the physical value of $k_F$ should be larger
than the expected finite quantum corrections
of order $\al c_{\mu\nu}$,
where $\al$ is the fine-structure constant.
If $k_F$ is 
eventually bounded to about $10^{-27}$
as estimated in subsection IV D,
then this would suggest 
the components of $c_{\mu\nu}$
might be expected to be smaller than about $10^{-25}$.

As in the CPT-odd case, 
the momentum-dependent radiatively induced corrections
in Eq.\ \rf{ABCDresult}
are irrelevant in the context of cosmological birefringence.
However,
these radiative corrections do modify the off-shell propagator
and might therefore be expected to generate small effects
under suitable circumstances.
For example,
there may be contributions to  
electromagnetic scattering cross sections,
governed by the ratio ${\al m_e}/M_P\approx 10^{-25}$.
Similarly,
a small correction to the Coulomb law might appear.
These issues lie beyond the scope of the present work.

In addition to the $c_{\mu\nu}$-linear one-loop contribution
obtained above,
there are also $c_{\mu\nu}$-linear higher-loop corrections 
in the extended QED.
The general structure of the contributions at any loop order
is given by Eq.\ \rf{gi}.
Although the detailed form of the scalar functions
$A$, $B$, $C$, and $D$ will differ,
the physically relevant corrections should also  
depend on $p^2 / m^2$,
since any terms independent of $p$ 
are expected to be absorbed by the renormalization procedure.
The above conclusions about cosmological birefringence
are therefore likely to remain valid.
Effects from higher-order Lorentz violation should also arise
in the extended QED but are probably of a size that is  
physically irrelevant.
Much of the above discussion should also hold for 
radiative effects in the full standard-model extension.
In this context,
note that off-diagonal terms in generation space 
cannot contribute at leading order.

\section{SUMMARY}

In this paper,
we presented a general Lorentz-violating extension 
of the minimal SU(3) $\times$ SU(2)$ \times$ U(1) standard model
including both CPT-even and CPT-odd terms,
and we discussed some of its theoretical and experimental properties.
The analysis was performed within the context
of a framework previously described
\cite{cksm},
which is based on spontaneous Lorentz and CPT violation 
occurring in an underlying theory of nature.

Despite the existence of terms causing
a certain type of Lorentz breaking,
the resulting theory preserves various desirable features
of standard quantum field theories
such as gauge invariance,
energy-momentum conservation,
observer Lorentz invariance,
hermiticity,
the validity of conventional quantization methods,
and power-counting renormalizability.
Other important features such as 
positivity of the energy,
microcausality,
and the usual anomaly cancellation are also expected.
We have demonstrated that the
usual breaking of SU(2) $\times$ U(1) symmetry
to the electromagnetic U(1) is maintained,
although the expectation value of the Higgs is slightly changed 
and the $Z^0$ field acquires a small expectation value.
The theory presented here 
appears at present to be the sole candidate
for a consistent extension of the standard model
providing a microscopic theory of Lorentz violation.

We have extracted extensions of 
several of the conventional varieties of QED
by considering limiting cases 
of the standard-model extension.
Part of the motivation for investigating extended QED
is the existence of high-precision tests 
of Lorentz and CPT invariance that involve electrodynamics.
A summary was provided of some recent studies
of possible experimental constraints.

Another major focus of this work
is the effect of the Lorentz violations 
on the photon sector.
The general pure-photon lagrangian can be written in a form
containing only two additional terms,
one CPT odd and one CPT even.
This lagrangian and the associated energy-momentum tensor
were discussed,
and it was found that the CPT-even component
has positive conserved energy
but that in the absence of a photon mass
the CPT-odd component can generate negative contributions
to the energy.
Despite this theoretical difficulty,
the two terms were retained for the whole analysis
so that the implications
for the full quantum theory could be examined.

The equations of motion
generalizing the Maxwell equations
in the presence of Lorentz violation
were obtained,
and their solution was outlined 
both using potentials and using fields.
Some technical complications arise relative to 
the case of conventional electrodynamics
\it in vacuo, \rm
but they can largely be overcome.
A key feature is that,
although there are still two independent propagating
degrees of freedom,
in the typical situation
the two modes obey different dispersion relations.
This implies a variety of interesting effects,
including birefringence in the vacuum.
We presented a few quantitative analogies with crystal optics
and showed that the presence of Lorentz violation 
means that the vacuum 
as experienced by an electromagnetic wave behaves like 
an optically anisotropic and gyrotropic transparent crystal 
exhibiting spatial dispersion of the axes.

A variety of terrestrial, astrophysical and cosmological
bounds on photon properties are known.
Sharp experimental limits on the photon sector
of the extended QED can be obtained
from the absence of birefringence on cosmological scales.
It has been shown in earlier work
\cite{cfj}
that the problematic CPT-odd term
is experimentally limited to scales
comparable to the Hubble length.

A significant result of this paper
is that most of the components of the CPT-even term 
could also be bounded experimentally 
from cosmological birefringence
with existing techniques.
This case is particularly interesting
as it has no evident theoretical difficulties
and appears to have been overlooked in
the previous literature.
Also,
unlike the CPT-odd term,
the CPT-even contribution exhibits a dependence on wavelength
that might provide a useful signature of the effect.
We have crudely estimated 
the attainable bounds,
which would be sensitive to suppressed Lorentz violation
in the general range considered here.

The paper also contains
a series of consistency checks on the theory,
primarily at the level of one-loop radiative corrections.
We discussed the cancellation of the various 
conventional anomalies in the standard-model extension
and considered other anomaly cancellations
that might occur in the underlying theory.
The latter were used to obtain a constraint 
on a set of coupling coefficients for Lorentz violation
in the standard-model extension.

We have investigated the feasibility of imposing 
tree-level vanishing of the problematic CPT-odd terms,
in light of possible radiative corrections that
could be induced from the non-photon sectors in the 
standard-model extension.
We have shown that the radiative corrections at
one loop are finite,
so it is unnecessary at this level of renormalization
to include a CPT-odd term in the original theory.
The finite corrections are gauge invariant but ambiguous,
a situation somewhat reminiscent of 
the usual anomaly calculations.
However,
if the theory underlying
the standard-model extension is anomaly free,
the CPT-odd effects in the photon sector
can be neglected at this level.
Generalizations of this argument may 
apply at higher loops.

For the CPT-even sector,
we have demonstrated by explicit one-loop calculation
that divergent radiative corrections appear.
A term of this type therefore must be present
in the original theory.
An experimental search 
for the associated renormalized coupling based,
for example,
on cosmological birefringence could be performed.
It is remarkable that physics 
associated with the Planck scale
might produce observable effects 
in measurements made at the largest scales in the Universe.

\acknowledgments
We thank R. Bluhm and P. Kronberg for discussions.
This work was supported in part
by the United States Department of Energy 
under grant number DE-FG02-91ER40661.

\end{multicols}

\begin{references}
\bibitem{fn1}
Note that the definitions of these terms  
differ slightly from those in Ref.\ \cite{cksm}.
\end{references}
\end{document}